\input{psfig.sty}
\documentstyle[12pt]{article}
\newfont{\feff}{cmti10}
\begin{document}

\title{Probability density and  scaling exponents of the moments of
longitudinal  velocity difference  in strong turbulence }

\author{ Victor Yakhot\\
Program in Applied and Computational Mathematics\\
Princeton University}

\maketitle

${\bf Abstract}.$
\noindent

We consider a few cases of homogeneous and
 isotropic turbulence differing by the mechanisms of turbulence generation.
The advective terms in the Navier-Stokes and  
Burgers equations are similar.  It is proposed  that the longitudinal
 structure functions $S_{n}(r)$
in homogeneous and isotropic three- dimensional
 turbulence are governed 
by a  one-dimensional 
equation of motion, resembling the 1D-Burgers equation,  with the strongly 
non-local pressure contributions  accounted for by 
galilean-invariance-breaking terms. 
The resulting equations, not involving  parameters taken 
from experimental data,
 give both 
scaling exponents and   amplitudes of the structure functions in an
 excellent agreement with experimental data. The derived probability 
density function  
$P(\Delta u,r)\neq P(-\Delta u,r)$
 but  $P(\Delta u,r)=P(-\Delta u,-r)$, in accord with the symmetry
 properties of the 
Navier-Stokes equations.
 With decrease
of the displacement $r$,  the  probability 
density, which   cannot be represented in a  scale-invariant form,
  shows smooth variation from the gaussian 
at the large scales to close-to-exponential function, thus demonstrating 
onset of   
small-scale intermittency. It is shown that accounting for
 the  sub-dominant 
contributions
 to the structure functions $S_{n}(r)\propto r^{\xi_{n}}$  
is crucial for derivation of the amplitudes of the moments of 
velocity difference.

\newpage

\noindent {\bf Introduction}
\\
Intermittency of turbulence,  not contained
in  the Kolmogorov theory, 
is one of the most intriguing and mysterious phenomena of 
continuum mechanics. Experimentally detected in the early sixties,  this
feature of high Reynolds number turbulent flows 
still remains  a major challenge
to turbulence theory.
Landau's 1942 remark [1] that the large- scale  fluctuations of
 turbulence production 
in the energy-containing range can invalidate the Kolmogorov theory
 was one of the motivations for  construction of various  ``cascade'' 
models attempting to explain
this  phenomenon,   manifested in
 the anomalous scaling of the structure functions 
$S_{n}(r)=<(u(x+r)-u(x))^{n}>\equiv <U^{n}>\propto
A_{n}r^{\xi_{n}}$ with the exponents $\xi_{n}\neq n/3$. The first model
 of this kind 
was proposed by Kolmogorov himself in 1962 [2]. Recently, some important 
analytic advances, leading to  evaluation of both scaling exponents 
$\xi_{n}$ and the amplitudes $A_{n}$ were made  for the problems of 
the passive scalar,  advected by a  random velocity field and 
the random-force-driven Burgers 
turbulence [3]-[7].
First, it was proposed by Kraichnan [3] that scalar structure functions 
$<(T(x)-T(x+r))^{2n}>$ can be solutions of homogeneous differential equations
, thus leading to the  non-trivial
values
of the exponents $\xi_{n}$ which could not be found from dimensional
 considerations. 
Then, Gawedzki and Kupiainen [4],  Chertkov  et. al.  [5]-[6] and 
Shraiman and Siggia [7] showed that, indeed,
it was the zero modes which were responsible for the anomalous scaling in
some limiting cases of the passive scalar problem. Similar results 
were arrived at in many of the following studies [8]-[10].

Polyakov's theory of the large-scale-random-force-driven
Burgers turbulence [11] was based on the assumption that 
weak small-scale  velocity fluctuations 
 ($|u(x+r)-u(x)|<<u_{rms}$ and $r<<L$,  
where $L$ is the the energy input-
 scale), obey the  galilean invariant dynamic equations,
 meaning that the integral scale and the random-force- induced
single-point $u_{rms}$ cannot enter the resulting expression for the 
probability density having the scale-invariant form:

$$P(U,r)\approx \frac{1}{r}F(\frac{U}{r})\eqno(1)$$

\noindent When $U\geq u_{rms}$   galilean invariance (GI),  even of
 the small-scale dynamics 
can be 
 violated and the PDF
scales with $U/u_{rms}$,  leading to  saturation of the scaling 
exponents $\xi_{n}=1$  for  $n>n_{c}=1$. This general
feature of   Burgers turbulence 
was confirmed  by   numerical experiments in which turbulence was
 generated by
different  random forces generating  various scaling exponents 
$\xi_{n\leq n_{c}}$ [12]-[14]. It was shown in Ref. [12]-[14] that 
 the value of the critical moment number
$n_{c}$ depended on the forcing function spectrum. Still,  at $n>n_{c}$
all $S_{n}\propto r$ indicating that decorrelation introduced by the noise
was too weak to prevent the shock formation. Recently Chertkov et. al. [15]
obtained a similar 
structure of the theory
considering a one-dimensional problem of a passive scalar advected by a
random velocity field. In the limit $r\rightarrow 0$ their PDF 

$$P(U,r)\propto \frac{1}{r}\frac{e^{-\frac{U^{2}}{u^{2}_{rms}}}}{U^{2}+r^{2}}$$

\noindent where, to simplify notation, we have set the values of all 
numerical constants equal to unity. Similar  result was also obtained in the 
work on the large-scale- driven Burgers turbulence in when space
dimensionality $D\rightarrow \infty$ by Bouchuod et. al.  [16].

\noindent Polyakov's idea [11] about the role of 
violation of  galilean  invariance  in
generation of  anomalous scaling resonates  with Landau's remark about
important influence  of the large-scale fluctuations  on 
the small-scale dynamics [1]. In  this paper we will 
attempt to combine the zero-mode and   the  GI-breakdown 
ideologies to derive  equations  governing the
 probability density function
 of the longitudinal velocity differences in strong turbulence.
 We would like to reiterate that  two measurements in the same 
 turbulent flow
performed in the laboratory and in the moving frame of reference
 (train or ship) must give the same answers. The ``violation of  galilean 
invariance''  is understood  hereafter only in a limited Polyakov's sense
as a possibility of $u_{rms}$ entering the probability density of
velocity difference. It will be shown that the details of the large-scale 
turbulence production mechanism are important,  leading to the
 non-universality of 
probability density function (PDF) of velocity difference. The
results will be compared with  experimental data. 

\noindent {\bf Formulation of the problem}
\\
Turbulence in Nature results from hydrodynamic instabilities of various 
laminar low-Reynolds-number flows.  The transition
 phenomena are 
is not universal,  depending 
on geometry, external fields etc. and,  at the present time, 
 cannot be
 accounted  for by
turbulence theory.  Hoping for some universality of the small-scale velocity
fluctuations in the inertial range, it is customary to develop a 
theory of turbulence, driven at the
 large scales by some terms  in the Navier-Stokes equations, one can treat 
theoretically. Usually, these large-scale forcing terms are assumed to be 
irrelevant in the inertial range.
Below we  discuss three models corresponding to  different
mechanisms of turbulence generation. First, let us consider 
the Navier-Stokes equations  on an infinite domain:

$${\bf v}_{t}+{\bf v\cdot \nabla v}=
{\bf f}-{\bf \nabla}p +\nu \nabla^{2}{\bf v}\eqno(2)$$

\noindent 

$${\bf \nabla\cdot v}=0$$

\noindent 
The gaussian large-scale forcing ${\bf f}$ is defined by the
 two-point correlation
 function 

$$<f_{i}({\bf x},t)f_{j}({\bf x'},t')>=P_{ij}
\kappa(|{\bf x-x'}|)\delta(t-t')\eqno(3)$$

\noindent where the projection operator $P_{ij}$ ensures the
 incompressibility
of the solution.  The force is assumed acting in the interval of
 wave-numbers 
$1/\Omega^{\frac{1}{3}}<<k_{0}\approx 1/L>>k_{d}$ where
 $\Omega\rightarrow \infty$ and 
$\eta=1/k_{d}$
are the volume of the system and dissipation scale, respectively. In other
words, the forcing spectrum is assumed  to decrease very rapidly outside the
 interval
$k\approx k_{0}$. In the limit $k<<k_{0}$ the system is in
 thermodynamic equilibrium and 
is described by the gaussian statistics and  energy spectrum 
$E(k)\propto k^{2}$ [17]. 
Thus, the order of the limits is: first we set the large value of
 $k_{0}$ and then 
$\nu\rightarrow 0$, so that
$k_{d}/k_{0}\rightarrow \infty$. 

In the case of the $\delta$-correlated forcing function (3) the source-related
contribution to the equation for the two-point correlation function
 can be written in a very simple way:

$$W=<v_{i}(x,t)f_{i}(x',t)>\propto \kappa(|x-x'|)$$

\noindent 

2.~Often, in the  real- life experimental situations, when
turbulence is generated in
 the vicinity of the boundaries (wall flows), nozzles (jets), bodies 
(wakes) and later on  transported into the bulk of the flow where
 the measurements take place, 
the model (1)-(3) is not  valid.  In this case a  better description is 
given
by the initial
 value problem,  taking  into account the
turbulence decay during this delay time. This will be also
 discussed in what follows.

3. In the third class of the flows turbulence is produced by a 
large-scale shear. Then,
introducing the so called Reynolds decomposition

$${\bf U=V+v}$$

\noindent where ${\bf v}$ ($<{\bf v}>=0$) is the fluctuation 
from the time-independent 
mean velocity $<{\bf U}>\equiv {\bf V}$, 
the equation for  velocity fluctuations 
${\bf v}$ is given by (1) with ${\bf f=0}$ and 
the turbulence  production term in the right side [1]:

$${\bf (v\cdot \nabla) V}$$

The force-free Navier-Stokes equations 
 are invariant under rotations, space-time translations,
parity and scaling transformations. They are also invariant under galilean
 transformations 
${\bf x}\rightarrow {\bf x+Vt}$ and 
${\bf v}\rightarrow {\bf v+V}$ where ${\bf V}$ is the constant
velocity vector of the moving frame. Boundary conditions and forcing can
 violate some or all of the
 symmetries of the equations (1). It is,  however, usually assumed that
 in the high Reynolds number 
flows with $\nu\rightarrow 0$ all
symmetries of the Navier-Stokes (even Euler) equations are restored in the limit $r\rightarrow 0$
and $r>>\eta$ where $\eta$ is the dissipation scale where the 
viscous effects become important. This means,
among other consequences,  
that in this limit the root-mean square velocity fluctuations $u_{rms}=\sqrt<{v^{2}>}$, not invariant
 under the
 constant shift, cannot enter the relations describing moments of velocity differences.
 If all this is
 correct, then the effective equations for the inertial-range 
velocity correlation functions must have the symmetries
 of the original Euler equations. For many
years this  
assumption was the basis of all turbulence theories. Based on the recent
 understanding of 
 Burgers turbulence [11]-[14],  some of the constraints on the 
allowed turbulence theories
 will
 be relaxed in what follows.

We are interested in the multi-point velocity correlation functions:

$$C_{n}({\bf x_{1},x_{2},....x_{n}})=<v_{i1}({\bf x_{1}})v_{i2}({\bf x_{2}})....v_{in}({\bf x_{n}})>$$

\noindent and longitudinal structure functions:

$$S_{n}(r)=<((u(x+r)-u(x))^{n}>\equiv <(\Delta u)^{n}>$$

\noindent where $u(x)$ is the $x$-component of the three-dimensional velocity field and
$r$ is the displacement in the direction of the $x$ axis.

In 1941 Kolmogorov, considering decaying turbulence,
 derived an  equation for $S_{3}(r)$ valid when $r\rightarrow 0$:

$$\frac{1}{r^{4}}\frac{\partial r^{4}~S_{3}(r)}{\partial r}=
-4{\cal E}+\frac{6\nu}{r^{4}}\frac{\partial ^{2} S_{2}(r)}{\partial r^{2}}
\eqno(4)$$

\noindent leading to the famous Kolmogorov $\frac{4}{5}$-law: in the limit 
$\nu\rightarrow 0$ and at $L>>r>>\eta$ where $\eta$ is the dissipation scale of turbulence 
defined as:

$$\nu S_{rr}(\eta)=O(1)$$

\noindent The third-order structure function in the inertial range 
$L>>r>>\eta$ is given by a $4/5$ law::

$$S_{3}(r)=-\frac{4}{5}{\cal E}r$$

In fact, the expression (4) is an approximation, neglecting  small 
contributions from time-derivatives and source functions, valid in the limit 
$r\rightarrow 0$. In general, it  must be modified to 
include the forcing function:

$$S_{3}(r)=
-\frac{4}{5}{\cal E}r
+O(r\overline{(\Delta u)(\Delta f)})\eqno(5)$$

\noindent We can see that for a white-in-time large- scale forcing function
$\kappa(r)=\kappa(0)-\gamma r^{2}$,
the sub-leading contribution to Kolmogorov $4/5$-law in the 
inertial range $r\rightarrow 0$ is $O(r^{3})$. In a more general situation
evaluation of the correction is not so simple.

In case of decaying turbulence  the sub-leading contribution to (5) is :

$$O(r\frac{\partial S_{2}(r)}{\partial t})\eqno(6)$$

\noindent while when turbulence is produced by the large-scale shear,
it is somewhat different  [18]:

$$O({\bf r\frac{\partial V}{\partial x}}S_{2}(r))$$

\noindent When $r$ is small, these terms can be neglected. However,
as will be shown below, they must be accounted for since the procedure 
of evaluation of the probability density $P(\Delta u,r)$ 
involves matching of the inertial range and large-scale solutions.

Derivation of (4)-(6) is based on the fact that due to 
the incompressibility condition, all transverse correlation functions can
 be expressed in terms
 of the longitudinal ones leading to the closed equations.  One can say
 that in a  very limited sense the procedure projects 
the original  three-dimensional problem onto a one-dimensional one.
 Regretfully,  due to the coupling between
 different components
 of the velocity field,   caused by the pressure terms, we cannot
 rigorously derive similar
 expressions for the high-order moments $S_{n}(r)$. The second
 difficulty is 
in the presence  of the dissipation anomaly (see below). Still, we can attempt to use
 some general
 features of the 
equations of motion and derive the scaling properties and  general form of 
the probability density function $P(\Delta u,r)$.

\noindent{\bf Equations for the probability density}
\\
One can introduce  a generating function

$$Z=<e^{\sum_{i} \lambda_{i}\cdot {\bf v(x_{i})}}>$$

\noindent where the vectors  ${\bf x_{i}}$ define  positions of the points 
denoted by the numbers $1<i<N$. Using the incompressibility condition,
the equation for $Z$ can be formally written:

$$\frac{\partial Z}{\partial t}+\frac{\partial^{2} Z
}{\partial \lambda_{i,\gamma}\partial x_{i,\gamma}}=I_{f}+I_{p}+D\eqno(7)$$

\noindent where $I_{f}$, $I_{p}$ and $D$ are related to the forcing,
 pressure and dissipation contributions to the Navier-Stokes equation
 (see below).
Although the advection contributions are accurately accounted for in
this equation, it  is 
not closed due to the  pressure and  dissipation terms.
The  latter 
can be treated using  Polyakov's  operator product expansion ideas [11],
while the 
former presents an additional difficulty to be dealt with. In what follows
we will be mainly 
interested in the moments of the two-point velocity differences which
in 
homogeneous and isotropic turbulence can depend only on 
 the absolute values of two vectors
(velocity difference ${\bf v(x')-v(x)}$ and displacement 
${\bf r\equiv x'-x}$) and the angle $\theta$ between them with $\theta=\pi/2$
and $\theta=0$ corresponding to transverse and longitudinal structure 
functions, respectively. In  spherical coordinates
the explicitly written advective terms in the equation (7)
involve

$$O(\frac{\partial^{2} Z}{\partial \lambda \partial r});~~O(\frac{1}{r}
\frac{\partial Z}{\partial \lambda});~~O(\frac{1}{\lambda}\frac{\partial Z}{\partial r});~~O(\frac{Z}{\lambda r})\eqno(8)$$

\noindent and various  trigonometric functions and angular differentiations. 
The theory of the longitudinal structure
functions,  presented below, is based on 
the assumption, correct for the third order moment $S_{3}(r)$ (see (4)),
that the angular dependence can be accounted for in a simple way and,
 as a consequence, 
 there exist an equation for $\theta=0$. This assumption is supported
by the following observations. It is easy to show that 
in the inertial range the second-order
structure function 

$$S_{2}(r,\theta)= \frac{2+\xi_{2}}{2}D_{LL}(r)(1-\frac{\xi_{2}}{2+\xi_{2}}cos^{2}(\theta))$$

\noindent with $D_{LL}(r)=<(u(x)-u(x+r))^{2}>$.
 More involved relation can
 be written for the fourth-order moment
[16]:

$$S_{4}(r,\theta)=D_{LLLL}(r)cos^{4}(\theta)-3D_{LLNN}(r)sin^{2}(2\theta)+
D_{NNNN}(r)sin^{2}(\theta)$$

\noindent where $D_{LLNN}=<(v(x)-v(x+r))^{2}(u(x)-u(x+r))^{2}>$
and $v$ and $u$ are the components of the velocity field perpendicular
and parallel to the $x$-axis, respectively.
 As one can easily deduce from the angular dependence, the functions 
 $D_{LLLL}(r)$ and $D_{NNNN}(r)$ denote
longitudinal and transverse structure functions, respectively. In the limit 
$\theta\rightarrow 0$

$$S_{4}(r,\theta)\approx D_{LLLL}(r)cos^{4}(\theta)+O(\theta^{2})$$

\noindent rapidly approaching $S_{4}(r,\theta=0)=D_{LLLL}(r)\equiv S_{4}(r)$.
Based on the above expressions,  we conclude that, as in the theory
 of multidimensional Burgers turbulence [16], where
the probability density of velocity difference can be represented as:
$P(U,r,\theta)\approx P(U,rcos(\theta))$,  
 here in the limit $\theta\rightarrow 0$ the mixing 
of the longitudinal and transverse correlation functions is very weak 
($O(\theta^{2})$). As a consequence, we assume that the closed equation 
for the probability density of longitudinal velocity differences exist.
Generalization of the theory
to the case of an arbitrary (not small) angle $\theta$ is the subject of 
an  ongoing study.
\\
We select  $N$  points ${\bf x_{i}}$ with $0<i<N$  on the $x$-axis and 
introduce the longitudinal generating function for the $N$-point 
correlation function:

$$Z_{N}=<e^{\lambda_{i}u(x_{i})}>\eqno(9)$$

\noindent where $\lambda_{j}=ik_{j}$. 
The equation of motion for $Z_{N}$ can be formally derived
(we neglect the subscript $N$ in what follows):

$$Z_{t}=\sum_{j} \lambda_{j}<e^{\lambda_{i}u(x_{i})}u_{t}(x_{j})>$$

Substituting the N-S equations into this relation gives:

$$Z_{t}=\sum_{j}\lambda_{j}<e^{\lambda_{i}u(x_{i})}[-u(x_{j})\frac{\partial u(x_{j})}{\partial x_{j}}+
f_{x}(x_{j})]>
+I_{T}+I_{p}+D\eqno(10)$$

\noindent where $f_{x}(x_{j})$ is the $x$-component of the forcing,
 $x_{j}$ is the coordinate of the $j^{th}$ point and 
$\frac{\partial}{\partial x_{j}}$ is the partial derivative in the $x$
 direction only. The summation in the expression (10) is over the positions
$x_{j}$ and over the Greek subscripts $\alpha=1;~2$ denoting the components
 of the velocity field in the directions perpendicular to the $x$-axis.  
The life-threatening  terms in (10) are: 

$$I_{T}=\sum_{j} \lambda_{j}<e^{\lambda_{i}u(x_{i})}[-v_{\alpha}(x_{j})\frac{\partial u(x_{j})}{\partial x_{\alpha j}}]>\eqno(11)$$

$$I_{p}=-\sum_{j}\lambda_{j}<e^{\lambda_{i}u(x_{i})}\frac{\partial p(x_{j})}{\partial x_{j}}>\eqno(12)$$

\noindent and  

$$D=\nu \sum_{j} \lambda_{j}<e^{\lambda_{i}u(x_{i})}\frac{\partial^{2} u(x_{j})}
{\partial x_{j}^{2}}>\eqno(13)$$

The theoretical and numerical work [16] on  the multi-dimensional Burgers equation 
led to the probability density and  moments of velocity difference  basically 
independent on the space dimensionality: the moments $S_{n\leq 1}(r)\propto r^{n}$
while $S_{n\geq 1}\propto r$. 
This is an indication that the shock production,
dominated by the 
longitudinal components of the nonlinearity $u_{i}\partial_{i} u_{i}$ 
(no summation over the subscript $i$),  prevails over the processes coming 
from the mixed  terms of the kind $u_{j}\partial_{j} u_{i}$ which can be neglected. 
In other words 
the multi-dimensional Burgers equation is well  approximated by the system 
of weakly interacting 1D-equations acting along various coordinate axis. 
It is 
clear that geometry of the objects generated by this system is very complex.
The recent paper  by Gurarie and Migdal [16], dealing with the two- and
 three-dimensional Burgers turbulence, introduced   an angle $\theta$ 
between  velocity difference and 
displacement vectors ${\bf v(x+r)-v(x)}$ and ${\bf r}$,
respectively,  and  using the instanton formulation, derived an
expression for the generating 
 function (see below) $Z_{2}$ in the form
$Z_{2}\propto exp((\lambda r)^{\gamma}f(cos \theta))$ with 
$\gamma=3/2$ independently on space dimensionality and 
$\lambda=|{\bf \lambda}|$ and $r=|{\bf r}|$. The calculated function 
$f(cos (\theta))$ ensured correct angular dependence of 
multi-dimensional structure functions. When $\theta=0$, 
the derived expression basically recovered the one-dimensional 
Polyakov's result.
 This result tells us that it is the projection of the velocity field 
on the direction of the displacement vector ${\bf r}$ that produces 
 dynamically significant contribution to the multi-dimensional
 structure functions and 
that the longitudinal structure functions in three-dimensional Burgers 
turbulence
are close to those  in the 1D-turbulence. Indeed, numerical 
simulations [16] of the three-dimensional Burgers turbulence 
revealed very complex velocity field  with the structure functions $S_{n}(r)$
very close
 to the ones, previously obtained in 
one-dimensional simulations. 
The possible smallness of  interaction between different components 
of  the advective terms is only part of the story.
The pressure contribution $I_{p}$ leads to 
effective  energy redistribution between components of the velocity field
and plays an important part in the Navier-Stokes dynamics. 
The pressure effects are nonlocal, instantaneously transporting  
information between different, even very remote,  parts of the flow. That is
 why 
 the effects coming from the 
boundary conditions and large scale forces cannot be neglected 
even in description of the small-scale phenomena.
The possible spontaneous 
breakdown of galilean invariance is the central assumption of this work.

The equation (10) can be rewritten as:

 $$Z_{t}=\sum\lambda_{i}\lambda_{j}\kappa(|x_{i}-x_{j}|)Z-\sum_{j}[\frac{\partial ^{2} Z}{\partial \lambda_{j}\partial x_{j}}-
\frac{1}{\lambda_{j}}\frac{\partial  Z}{\partial x_{j}}]
+I_{T}+I_{p}+D\eqno(14)$$

\noindent where the large-scale gaussian random force 
is defined  in the limit $r\rightarrow 0$ by the correlation function:
$\kappa(r)=\kappa(0)- \gamma r^{2}$.
Approaching the integral scale $L$ the force correlation function 
$\kappa(r)$ rapidly goes to zero. We will see that the equation for the 
probability density of the velocity difference $P(U,r)$ where $U=u(x+r)-u(x)$
contains the combination $\kappa(0)-\kappa(r)$ which is large at the large
 scales $r\rightarrow L$. That is why the large-scale dynamics, dominated
by the forcing term, show close to  gaussian behavior of at least the first 
few moments $S_{n}$.

Evaluation of $I_{T}$, $I_{p}$ and $D$ in (14)  is a difficult
problem and we have to make some assumptions.
It is seen from the definition of the generating function $Z_{N}$  that
when two points merge, i.e. $x_{i}\rightarrow x_{j}$ the $N$-point 
generating  function becomes the $N-1$-point generating  function with
$\lambda=\lambda_{i}+\lambda_{j}$. This means that if, for example,
 the equation for $Z_{2}$ contains the two-point sum:

$$\varphi(\lambda_{1},u(x_{1}),x_{1},\frac{\partial}{\partial x_{1}})Z_{2} +
\varphi(\lambda_{2},u(x_{2}),x_{2},\frac{\partial}{\partial x_{2}})Z_{2}
 \rightarrow
a(\lambda, y)\varphi(\lambda,u(x_{1}),x_{1},\frac{\partial}{\partial x_{1}})Z_{1}$$

\noindent Here $\varphi$ depends on the structure of the equation of motion
and 
$y=|x_{i}-x_{j}|\rightarrow 0$.
We assume that in this limit the unknown function $a(\lambda,0)$ is finite. 
Not all functions $\varphi$ satisfy this equation. 
For example:
$\varphi=\lambda$
and $\varphi=\lambda\frac{\partial }{\partial \lambda}$,  
do. The functions $\varphi$ can also include  space derivatives.
Combined with the general 
symmetry properties of (14) we can narrow the  class
of possible solutions and derive equation for the probability density.
It follows from the Navier-Stokes equations that the theory must be invariant under transformation:
$\lambda\rightarrow -\lambda$ and $x_{j}\rightarrow -x_{j}$.
In what follows we will  adopt  Polyakov's result  that the main  effect of
 the longitudinal
 part of the dissipation term $D$ is a renormalization of the coefficient in
 front of 
the $O(1/\lambda_{j})$ terms in the right side of (14). 
Based on the above considerations,  we  can write two   models
for the $N$-point generation function, corresponding to the  different
mechanisms of turbulence production introduced above (see below):

 $$Z_{t}=\sum\lambda_{i}\lambda_{j}K(|x_{i}-x_{j}|)Z-\sum_{j}[\frac{\partial ^{2} Z}{\partial \lambda_{j}\partial x_{j}}-
\frac{b}{\lambda_{j}}\frac{\partial  Z}{\partial x_{j}}]
+C\sum \lambda_{j}\frac{\partial}{\partial \lambda_{j}}Z+I_{T}+I_{p}+D'\eqno(15)$$

\noindent for the cases  2 and 3 and

 $$Z_{t}=\sum\lambda_{i}\lambda_{j}\kappa(|x_{i}-x_{j}|)Z-\sum_{j}[\frac{\partial ^{2} Z}{\partial \lambda_{j}\partial x_{j}}-
\frac{b}{\lambda_{j}}\frac{\partial  Z}{\partial x_{j}}]+I_{T}+I_{p}+D'\eqno(16)$$

\noindent to describe turbulence generated by the white-in-time
forcing (3). Here   $D'$ involves only ``transverse'' components of 
the $D$-term
defined below. The $O(\lambda\frac{\partial Z}{\partial \lambda})$- term
 in (15)
comes from the $O({\bf v})$ turbulence production, while 
$K(r)=K(0)-\beta r^{2}$, leading to the negligibly small $O(r^{2})$
sub-leading contributions to the velocity difference structure functions,
ensures the close-to-gaussian single-point
probability density. The equations (15)-(16)   violate neither
``fusion rules''  introduced above nor  general symmetries of
 the Navier-Stokes equations.
 By dimensionality,  the coefficient $C/\kappa(0)=O(1/u_{rms}^{2})$.  The 
$O(\kappa(r))$ term in the 
equation (16), stems from the forcing function (3). 
Below we will discuss the two cases in detail.

If one is interested in a single-point probability density, the equation (15) is to be solved 
for all $x_{j}=x$ and $\lambda=\sum_{j} \lambda_{j}$. All space derivatives disappear due
to homogeneity of turbulence and we have an expected 
result [17]:

$$P(u)=\sqrt{\frac{2}{\pi}}e^{-\frac{u^{2}}{2u^{2}_{rms}}}$$

\noindent This fixes the value of the coefficient $C$. Thus, the equation (15) yields  
gaussian distribution of the single-point velocity field. This result will be used below as
a matching condition for the probability density $P(U,r)$ in the large-scale  limit
$r\rightarrow L$.  The experimentally observed 
single-point probability density $P(u)$ is very close to but not equal to
gaussian deviating from it
at the large values of velocity fluctuations $u>>u_{rms}$.
The theory, developed here, is applicable to any expression for $P(u)$,
not only to the  gaussian. Still, the gaussian, which will be used below to 
compare the theoretical  predictions with the data,  
is a very good approximation.

We need an  equation
 for the
the generating function $Z_{2}$ with $\lambda_{1}+\lambda_{2}=0$
giving:

$$Z_{2}(\lambda,r)=<e^{\lambda (u(x+r)-u(x))}\equiv <e^{\lambda U}>\eqno(17)$$

In a statistically steady state:

 $$2\frac{\partial ^{2} Z_{2}}{\partial \lambda \partial r}-
\frac{2b}{\lambda}\frac{\partial  Z_{2}}{\partial r}=
\lambda^{2}(K(0)-K(|x_{i}-x_{j}|))Z_{2}
+I_{T}+I_{p}+D'-
\frac{u_{rms}}{L} \lambda \frac{\partial Z_{2}}{\partial\lambda}
\eqno(18)$$

\noindent
Where 

$$I_{T}=\lambda<[v_{\alpha}(x_{2})\partial_{x_{2},\alpha}u(x_{2})-
v_{\alpha}(x_{1})\partial_{x_{1},\alpha}u(x_{1})]e^{\lambda(u(2)-u(1))}>\equiv\lambda
<i_{T}e^{\lambda U}>$$

$$I_{p}=<\lambda[\frac{\partial p(x_{2})}{\partial x_{2}}
-\frac{\partial p(x_{1})}{\partial x_{1}}]e^{\lambda(u(2)-u(1))}>\equiv\lambda <i_{p}e^{\lambda U}>$$

and

$$D'=\lambda\nu<[\partial^{2}_{x_{2}\alpha} u({\bf x_{2}})
-\partial^{2}_{x_{1}\alpha} u({\bf x_{1}})]e^{\lambda(u(2)-u(1))}>\equiv \lambda <d' e^{\lambda U}>\eqno(19)$$

The point  merging in  $I_{T}$ and $I_{p}$ 
must be regular while the same procedure in the dissipation term $D'$ involves
 divergences which are cancelled  by viscosity $\nu\rightarrow 0$. 
As was pointed out in  [11] the longitudinal,  $O(\partial_{x}^{2} u({\bf x}))$,
  components of the 
$D$-term result  in the renormalization of the coefficient in front of
the $O(1/\lambda)$ -  contribution to the right side of (14). 
We are still left
 with the remaining $O(\partial^{2}_{\alpha} u({\bf x}))$ -  piece of the dissipation anomaly,
 pressure terms $I_{p}$ and the $I_{T}$ - contributions,  mixing all components of the 
velocity field.
Having in mind  general fusion rules, considered above, and the fact that
the equation is invariant under transformation $\lambda\rightarrow -\lambda$
and $r\rightarrow -r$ we, not being concerned with preservation of galilean
 invariance, write the equation for $Z_{2}$  corresponding to (15):

 $$\frac{\partial ^{2} Z_{2}}{\partial \lambda\partial r}-
\frac{B^{0}}{\lambda}\frac{\partial  Z_{2}}{\partial r}
=
\frac{A}{r}\frac{\partial Z_{2}}{\partial \lambda}-\frac{u_{rms}}{L}
\lambda\frac{\partial Z_{2}}{\partial \lambda}\eqno(20)$$

\noindent 
The equation (20) includes the 
derived above expression for  
 the coefficient $C$ and 
the unknown parameters $B^{o}$ and $A$ to be determined
 from the theory. The characteristic time in (20) $T\approx L/u_{rms}=O(1)$
is independent on the displacement $r$. A natural
generalization  of this model is (20) with 
the last term in the right side:

$$\frac{1}{T(r)}\lambda\frac{\partial Z_{2}}{\partial \lambda}$$

\noindent with $T(r)\propto r^{\xi_{\tau}}$ with the exponent $\xi_{\tau}$
depending on the physics of the problem.  In Kolomogorov turbulence 
$\xi_{\tau}\approx 2/3$.

The model corresponding to (16) is:

 $$\frac{\partial ^{2} Z_{2}}{\partial \lambda\partial r}-
\frac{B^{0}}{\lambda}\frac{\partial  Z_{2}}{\partial r}
=
\frac{A}{r}\frac{\partial Z_{2}}{\partial \lambda} +\gamma r^{2}\lambda^{2}Z\eqno(21)$$

\noindent where the $O(r^{2})$- contribution is to be kept.
 The equations (20)-(21) are  based on the assumption
 that the dynamic role of the pressure and  dissipation terms 
 is in the renormalization
 of
 the coefficients in front of the already- present advective contributions (8)
to the equation (7). Similar 
assumption was 
fruitful  in the theory of Burgers turbulence [11].
Except for the last  terms in the right side,  these equations are 
the same as   the one in Polyakov's theory of  Burgers turbulence with 
the $B^{o}$ - term simply
 renormalizing part of the advection effects, present in the original
 equations. 
The meaning of the $A$-term  will be discussed  in detail  below.
It will be shown that  it is not only responsible for
  prevention of the shock
formation but it also makes the weak ($|U|<<u_{rms}$ ) structures
 of the Navier-Stokes dynamics much stronger than their counterparts
 of  Burgers turbulence.
It is clear that  due to homogeneity of 
turbulence,  all space derivatives and related to
them $A$-contributions  go to zero when $r\geq L$
. In the one-dimensional case the dissipation anomaly $D$, 
discussed in detail by Polyakov, leads only to a relatively small
 modification of a
corresponding coefficient . It will be shown  that in case of 
three-dimensional turbulence $B^{o}\approx -20$, part of 
 which is to be attributed to
 a large pressure
effects  preventing the shock formation.  In the resulting
 dynamic equations (20)-(21) the contributions
 from  $D$ and $I_{p}$ are mixed and the origins of each term,
hidden in numerical values of the coefficients $b$, $B^{o}$ and $A$,  are 
 not easy
to establish.
The equation (20), explicitly involving the single-point $u_{rms}$
is suited for description of the generating function in both
$r\rightarrow 0$ and $r\rightarrow L$ limits. In the inertial range,
 where the displacement
 $r$ is small,  the $O(u_{rms})$ and the forcing contributions  can be neglected. They are 
important, providing  the large-scale gaussian 
matching constraint,  needed  for determination of the amplitudes 
of the structure functions $S_{n}$. Thus, the equations (20)-(21),
describing the correlation functions in the inertial range,  differ by 
the last
in the right side terms reflecting the details
 of the large-scale turbulence generation processes. It will be shown below
that this difference is responsible for non-universality
of the probability density function of velocity difference.

In the limit $r\rightarrow 0$ the equation for the probability
 density is derived from (20)-(21) readily:
$$-\frac{\partial }{\partial U}U\frac{\partial P}{\partial r}-
B^{o}\frac{\partial P}{\partial r}=-\frac{A}{r}\frac{\partial}{\partial U}U~P +
\frac{u_{rms}}{L}\frac{\partial^{2}}{\partial U^{2}}UP\eqno(22)$$

\noindent 

$$-\frac{\partial }{\partial U}U\frac{\partial P}{\partial r}-
B^{o}\frac{\partial P}{\partial r}=-\frac{A}{r}\frac{\partial}{\partial U}U~P -
\gamma  r^{2}\frac{\partial^{3} Z}{\partial U^{3}}\eqno(23)$$

\noindent{\bf Properties of the solution}
\\ 
Multiplying (22) by $U^{n}$ and assuming existence of all moments leads to:

$$\frac{\partial S_{n}}{\partial r}=
\frac{An}{n+B}\frac{S_{n}}{r}+
\frac{u_{rms}}{L}\frac{n(n-1)}{n+B}S_{n-1}(r)\eqno(24)$$

\noindent where  $B=-B^{o}>0$.  The equation (24) is to be solved under
 the constraint (4) which is the result of the energy conservation 
inherent to  the Navier-Stokes equations. This is the consequence  of the 
renormalization ideology leading to  (24),  which is a model,
not rigorously derived from (1)-(3), but based on some general symmetries
of the Navier-Stokes equations. That is why the $4/5$ - law comes out
of (24) only for a particular set of  parameters. In a ``final'' theory
the rigorous equation for $S_{n}(r)$ must automatically produce
the correct result for $S_{3}(r)$.
Neglecting the last term in the right side of (24) (see below) 
the solution for $S_{3}(r)$ 
in the limit $r\rightarrow 0$ is derived readily:

$$S_{3}=Nr^{\frac{3A}{3+B}}$$

\noindent This means that the coefficient 
 $A=\frac{3+B}{3}$ and $N=-\frac{4}{5}{\cal E}$. 
Seeking the solution in
the form $S_{n}\propto r^{\xi_{n}}$ we obtain:

$$\xi_{n}=\frac{(3+B)n}{3(n+B)}\eqno(25)$$

\noindent The normal Kolmogorov scaling $\xi_{n}=n/3$ , corresponding to 
the no - intermittency case
is achieved in the limit $B\rightarrow
\infty$. The maximum, Burgers-like, intermittency
  with all exponents $\xi_{n}=1$ due to the $tanh x$-shocks  
is recovered when  $B=0$.

\noindent Deriving (25) the contribution of the order 

$$\frac{u_{rms}}{L}\frac{n(n-1)S_{n-1}(r)}{n+B}$$

\noindent was neglected in comparison with the 
$O(\frac{nS_{n}}{(n+B)r})$ terms.
Substituting the expression for $S_{n}=A_{n}r^{\xi_{n}}$ with $\xi_{n}$
from (22)  gives a general solution:

$$S_{n}=A_{n}r^{\xi_{n}}+
\frac{u_{rms}}{L}A_{n-1}
\frac{n(n-1)}{n+B}\frac{r^{\xi_{n-1}+1}}{\xi_{n-1}-\xi_{n}+1}\eqno(26)$$

The coefficients  $A_{n}$ will be derived below. The expression (26) 
shows that due to the presence of the 
$O(r^{\xi_{n-1}+1})$ subdominant contributions, the experimental 
determination   of the scaling exponents 
is a difficult task and that proper accounting for it can lead to substantial
broadening of the inertial range and more accurate determination of
 the numerical values of the scaling exponents $\xi_{n}$. In addition,
it establishes the relation between the amplitudes of 
 the odd and even-order moments.

By definition of the integral scale, adopted in this work, the third-order 
moment $S_{3}(r=L)=0$. Since $u_{rms}\approx ({\cal E} L)^{\frac{1}{3}}$,
the expressions (25)-(26) give:

$$A_{3}=-\frac{4}{5}=-9A_{2}\frac{B+2}{(B+3)^{2}}$$

\noindent Taking in accord with the closures derived from various  
renormalized perturbation expansions $A_{2}\approx 2$ leads to $B\approx 20$ 
and:

$$\xi_{n}=\frac{23}{3}\frac{n}{n+20}\eqno(27)$$

The calculated values of the exponents $\xi_{1/5}=0.0759;~~
\xi_{1/4}=0.0946;~~\xi_{1/2}=0.187;~~
\xi_{1}=0.365;~~
\xi_{2}=0.696;~~\xi_{4}=1.278;~~
\xi_{5}=1.533;~~\xi_{6}=1.769;~~\xi_{7}=1.988;~~\xi_{8}=2.190;~~
\xi_{9}=2.379;~~\xi_{10}=2.555$ are indistinguishable from the best
 available experimental data. One has to keep in mind that the value of the 
parameter $B$ can be non-universal slightly varying from flow to flow. 
This can lead to some non-universality of the exponents. The
 comparison of the 
magnitudes  of the exponents,  given by (27),  with the outcome
 of numerical simulations by Chen [19]
is presented on Fig.1.  The expression (27) predicts  saturation of the values of the exponents 
$\xi_{n}$
at $\xi_{\infty}\approx 20/3$ in agreement with some general  ideas 
based on the path integral representation of the solution of the passive
 scalar problem Chertkov [20]. 

The expression for $S_{n}(r)$ corresponding to the model (23) is:

$$S_{n}=A_{n}r^{\xi_{n}}+\gamma  A_{n-3}\frac{n(n-1)(n-2)}{n+B}
\frac{r^{3+\xi_{n-3}}}{3+\xi_{n-3}-\xi_{n}}\eqno(28)$$

\noindent It follows from (28) that 

$$S_{3}(r)=-\frac{4}{5}r+O(r^{3})$$

\noindent in accord with an exact result.

To calculate the value of  parameter $B$ one has 
 to solve the equation 
for the probability density $P(U,r)>0$ 
going to  gaussian in the limit $r/L\rightarrow 1$.
The equation is:

$$\frac{\partial }{\partial U}U\frac{\partial P}{\partial r}-
B\frac{\partial P}{\partial r}=
\frac{1}{r}(\frac{3+B}{3})
\frac{\partial}{\partial U}U~P\eqno(29)$$

It may be somewhat easier to deal with the equation for the generating function $Z_{2}$:

 $$\frac{\partial ^{2} Z_{2}}{\partial \lambda\partial r}-
\frac{B}{\lambda}\frac{\partial  Z_{2}}{\partial r}=
\frac{1}{r}(\frac{3+B}{3}) \frac{\partial Z_{2}}{\partial \lambda}$$

The structure of the solution is clear from the scaling of the moments
 derived above:

$$Z_{2}=\sum_{0}^{\infty}(-1)^{n}A_{n}
\lambda^{n}r^{\frac{(3+B)n}{3(B+n)}}$$

\noindent with as yet unknown amplitudes  $A_{n}>0$ which will be
 evaluated below.  The most important
 outcome of this expression is the fact that the odd-order moments 
$S_{2n+1}<0$
which means that the PDF $P(U,r)\neq P(-U,r)$. It is clear that 
  $P(U,r)=P(-U,-r)$  in accord with 
 the symmetry 
of the Navier-Stokes equations. 

To evaluate  the probability density function we need to
  match the inertial range PDF 
with either 
the energy  containing or dissipation range probability density functions. 
Based on the result  
for the single-point PDF [17],  one has to seek   the solution to (25)
that becomes very close to gaussian 
at the scales larger than some integral scale $L$  
. This condition can serve as a 
definition of the integral 
scale. In reality, the integral scale is never much smaller than the size of the system.  The 
  experimental data show that at the large scales the PDF is  
close to  gaussian with some  small deviations  seen at the far tails where $U>>u_{rms}$.
Based on the data we can safely assume that at the large scales the firstt few (10-20)
moments are very close to their gaussian values. The non-zero value of the odd-order
 moments $S_{2n+1}(r)$ with $n\geq 1$ implies the asymmetry of $P(U,r)$. However, this 
asymmetry is very small with $S_{2n+1}(r)/s_{2n+1}(r)<<1$ where 
$s_{2n+1}(r)=<|u(x)-u(x+r)|^{2n+1}>$  is often measured by the experimentalists. 
It has been shown [21]  that up to $n\approx 5$ 
this ratio is in the range
 of $\approx 0.1$
and  that the experimentally observed PDF can be made 
symmetric by a  signal-filtering procedure,  leading to near-vanishing  of
the odd-order moments. The procedure left the even-order moments
 unchanged,  indicating
that the PDF asymmetry only weakly influences  the even-order moments.
The data will be presented below.  

Now we set $L=1$  and, neglecting all odd-order moments, 
calculate the PDF directly from 
the generating  function:

$$P_{e}(U,r)=\frac{2}{\pi}\int_{0}^{\infty}cos(kU)\phi(k,r)\eqno(30)$$

\noindent with

$$\phi(k,r)=
1+\sum_{n=1}^{\infty}(-1)^{n}A_{2}^{n}(2n-1)!!~r^{\xi_{n}}\frac{k^{2n}}{2n!}\eqno(31)$$

\noindent
  We see that this expression gives $S_{2n}=A_{2}^{n}(2n-1)!!~r^{\xi_{n}}$ 
leading to the 
desired gaussian values at $r=L=1$. The same result 
$P(U,r)\propto exp(-\frac{U^{2}}{2r^{\frac{2}{3}}})$ is recovered 
in the limit $B\rightarrow \infty$ . In the opposite limit $B\rightarrow 0$ 
 the probability density tends to 
 
$$P_{e}(U,r)=(1-r)\delta(U)~+~r\sqrt{\frac{2}{\pi}}e^{-\frac{U^{2}}{2 u_{rms}^{2}}}$$

\noindent giving all   $\xi_{n}=1$. This corresponds to
Burgers turbulence in the GI-broken  range [11]. 
This result is very close  to the outcome of  the theory of the Burgers
 turbulence 
in the limit of  space dimensionality $D\rightarrow \infty$ [16],
predicting $P(U,r)=(1-r)\delta(U-r) +r\psi(\frac{U-r}{u_{rms}})$.
This fact is an indication that the galilean-invariance-breaking
terms in the equations of motion, obtained in this work,  are quite close 
to the truth.  It is clear that
to reproduce the shifted $\delta$- function of Ref.[16]
we have to abandon the 
simplification of  treating 
the PDF as an even function of $U$.

To uncover the inner structure of the  $\delta$-function the data
 are to be presented in coordinates $U/r^{\alpha}$ with $\alpha\approx 0.365$
(see Figures below). 
We can  
compare the prediction
 given by (31)
with the experimental data on 

$$C_{2n}(r)=\frac{S_{2n}(r)}{S_{2}^{n}(r)}=(2n-1)!!~r^{\xi_{2n}-n\xi_{2}}\eqno(32)$$

\noindent where all $\xi_{n}$ are given by (22).
Sreenivasan et. al [21] measured $C_{2n}(r)$ in the 
 low Reynolds number experiment ($R_\lambda\approx 200$) 
 conducted in the laboratory 
boundary layer. The results of Ref. [21] for $r\approx 0.1-0.2$ are:

$$C_{4}\approx 3.6;~~ C_{6}\approx 24.8;~~ C_{8}\approx 261;~~ C_{10}\approx 3770$$

\noindent which are to be compared with our predictions $r=0.2$:

$$C_{4}=3.16;~~C_{6}=25.05;~~C_{8}=272;~~C_{10}=3256:$$

The intermittency grows  strongly with decrease of the displacement $r$. 
For example:
for $r=0.1$ the relations derived in the present paper give:

$$C_{4}=3.42;~~C_{6}=31.26;~~C_{8}=412;~~C_{10}=6184;$$

These numbers agree extremely well with the 
results of numerical simulations by Chen [22] 
who was able to produce the data set 
at $R_\lambda=200$ consisting of a few billion points. The somewhat lower
 values 
of $C_{10}$ obtained in the physical  experiments can be attributed to
the insufficient statistics of the real-life data set. Recent high Reynolds number
experiments ($R_{\lambda}\approx 15000$) [23] produced similar  results
 for the moments $S_{n}$ with $n<6$.
The comparison between theory and experiment is somewhat difficult due
 to an uncertainty in  the theoretically
needed ratio $r/L$ where $L$ is a poorly defined integral scale
of turbulence.
The inertial range prediction
prediction of this paper
$C_{4}\approx 3r^{-0.114}$ agrees  well with the data obtained in 
the high Reynolds number experiments ($R_{\lambda}\approx 1500-2500$)
in the planetary boundary layer [21].

The evaluated probability density functions are compared with
the outcome of the measurements of Noullez et. al.  [24]
 on the Figs.2-4.
The high quality experimental data on the transverse structure functions
were obtained using  real-space measurements
in the air jet.  Thus, for a time being, comparing theory with experiment
here  we assume that the probability density of the 
the longitudinal  velocity differences  has the same shape as the one    of 
transverse velocity differences. 
The quantitative agreement for $1/4<r/L\leq 1$ is very good. 
We were not able to plot
$P(U,r)$ for very small values of the displacement $r$ but the data on 
Fig.4 show the tendency of the PDF to the $\delta$-function
in the limit $r\rightarrow 0$ in accord with the analytic asymptotics.
The Fig.2 presents the same curves,
plotted in the coordinates $U/r^{1/3}$. One can see increasing deviations from
the gaussian at $r/L=1$ with decrease of the displacement $r$. It is clear from
 the Figure
that the probability density cannot be represented in the scale-invariant form
 (1). This is 
the manifestation of the intermittency.
The calculation was performed using $Mathematica^{TM}$
which failed to produce the generation function $\phi(k,r)$ with $k>5$.
The  information on  $\phi(k,r)$ with $0<k\leq 5$ was sufficient 
to calculate $P(U,r)$ with the accuracy $\approx 1-2 \%$ in the range
of variation $0<U<3-4$ and $1/4<r<1$. The fact that at  $r=1$ the PDF 
must be gaussian serves as a good test of the quality  of the 
numerical procedure.

\noindent Using the derived  expression for $Z_{2}$, one  can easily 
evaluate 
correct asymmetric probability 
density $P(U,r)$ giving the values of the 
odd-order moments in agreement
 with experimental data. All one has to do is to
introduce a term involving the $sin$-contribution to the Fourier-transform 
in (26)-(27), generating    non-zero odd-order moments. Demanding that 
$S_{2n+1}(L)=0$ we have from (26):

$$A_{2n+1}\approx -\frac{u_{rms}}{({\cal E}L)^{\frac{1}{3}}}\frac{2n(2n+1)A_{2n}}{(2n+1+B)(\xi_{2n}-\xi_{2n+1}+1)}$$

\noindent valid for $n>1$. Setting $u_{rms}
\approx ({\cal E}L)^{\frac{1}{3}}$ and taking 
$A_{2}\approx 2$ we obtain: 

$$A_{3}\approx -0.8;~~A_{5}\approx 23.0;~~ A_{7}\approx 650:~~$$

In the case of the large-scale- driven turbulence the expression for $A_{2n+1}$
is derived in a similar manner:

$$A_{2n+1}=-\gamma  A_{2n-2}\frac{2n(2n+1)(2n-1)}{(2n+1+B)
(3+\xi_{2n-2}-\xi_{2n+1})}$$

This relation contains two unknowns $\gamma$ and $B$. Assuming universality
of the exponents, implying $B=20$, we find  from the relation for $A_{3}=-4/5$
that
$\gamma\approx 6$. In the correct dimensional units $\gamma \approx 6{\cal E}$.
This will be important below for 
 the quantitative comparison of theoretical predictions 
with   experimental data. This result 
has some interesting consequences. 
Keeping the amplitudes $A_{2n}$ equal to the ones derived above:

$$A_{5}\approx 14;~~A_{7}\approx 373;~~A_{9}\approx 28500;~~$$

We see that even if the exponents are universal, the amplitudes of the moments
are not,  meaning that the shape of the probability density
 can vary from flow to flow. It is interesting that to experimentally
 observe this effect one has to measure high-order moments since the 
first few structure functions 
in different  flows,  considered here,  seem to be close. Hereafter we will
 be mainly
interested in the model (20) which is more closely related to  physical experiments.

The expression 

$$\phi_{1}(k,r)=\sum_{1}^{\infty} A_{2n+1}r^{\xi_{2n+1}}\frac{(-k)^{2n+1}}
{(2n+1)!}$$

is to be substituted into the integral:

$$P_{o}=\frac{1}{\pi}\int_{-\infty}^{\infty}sin(kU)\phi_{1}(k,r)~dk
\eqno(34)$$

\noindent to give a total asymmetric PDF $P(U,r)=P_{e}(U,r)+P_{o}(U,r)$.
A very accurate parametrization of the central  part of the PDF, 
corresponding to not-too-large values of $U$, illustrating appearence of
 the asymmetric PDF is: 

$$\phi_{1}(k,r)=-\frac{2}{15}rk^{3}\phi(k,r)\eqno(35)$$

\noindent giving:

$$P(U,r)=P_{e}(U,r)+P_{o}(U,r)=
P_{e}(U,r)+\frac{2r}{15}\frac{\partial^{3} P_{e}(U,r)}{\partial U^{3}}
\eqno(36)$$

\noindent This expression is 
approximate and serves  only to illustrate the mechanism 
of appearence of the experimentally observed asymmetric PDF $P(U,r)$.
 The central part of the probability density can can be very accurately parametrized  by 
the formula  (36) with:

$$P_{e}(U,r)=\frac{N}{r^{\alpha}}e^{-x\tanh\frac{x}{a}}\eqno(37)$$

\noindent where  $N$ is a normalization constant,
$x\approx \frac{U}{r^{\alpha}}$ with, as above,  $\alpha\approx 0.365$.
  The parameter $a\approx 2~-~4$.
The expression (31)-(32) gives a  very good approximation  for
 the moments $S_{n}(r)$ with $n\leq 10$. For example: 
$A_{2}=1.9-2.6;~A_{3}=-0.8;~
A_{4}=20-28;~A_{5}=-15.4-21;~A_{6}\approx 657;~A_{7}\approx -785;~$ etc., 
very close to the first few amplitudes $A_{n}$, calculated above. Figs. 5-6
show the probability density function $P(U,r)$ given by (36)-(37).

The theory can be approximately generalized to the  case
of  correct anomalous exponents:

$$R_{n}=\frac{S_{n}}{S_{2}^{\frac{n}{2}}}\approx 
\frac{A_{n}}{A_{2}^{\frac{n}{2}}}r^{\xi_{n}-\frac{n\xi_{2}}{2}}
\eqno(38)$$

\noindent The expression (38) gives $S_{3}$ in accord with the Kolmogorov relation.
For $r\approx 0.1$ and 
$A_{2}\approx 2 $ we derive:
$R_{3}=-0.31;~R_{5}=-4.2;$.

To assess  internal consistency of the  theory and understand the role and
 meaning 
of the $A$-contribution to (20),  we write a formally exact 
equation for the two-point generation function:

$$2\frac{\partial ^{2} Z_{2}}{\partial \lambda \partial r}-
\frac{2}{\lambda}\frac{\partial  Z_{2}}{\partial r}=
\lambda^{2}(\kappa(0)-\kappa(|x_{i}-x_{j}|))Z_{2}
+I_{T}+I_{p}+D 
\eqno(39)$$

The equation for the probability density $P(U,r)$ can also be formally written:

$$\frac{\partial }{\partial U}U\frac{\partial P}{\partial r}+
\frac{\partial P}{\partial r}=-\frac{1}{2}\frac{\partial^{2} }{\partial U^{2}}[T(U,r)P(U,r)]\eqno(40)$$

\noindent where

$$T(U,r)=<i_{p}+i_{T}+d|~U>\eqno(41)$$

\noindent is the conditional expectation value of $T=i_{p}+i_{T}+d$ for a 
fixed value of
velocity difference $U$. Here $d=d'+\nu (u_{x'x'}(x')-u_{xx}(x))$.
The formulation of probability density functions in terms of 
conditional dissipation and production was introduced in [25]-[26]
 and became 
a subject of intense
investigations.  Three-dimensional turbulence problem 
is somewhat different since
 there is no way one can separate  various contributions to $T(U,r)$. 
Indeed, we are
 dealing here
 with the projection of a three-dimensional dynamics onto a line. 
This leads to  
violation  of all 
conservation laws and it is clear that the pressure-induced 
processes of the correlation- redistribution 
between different components of velocity, probably conservative in the original
 equation of motion, can lead
 to the dissipation-like effects   of  $T(U,r)$. We can see from the above
 definitions
that due to homogeneity of the turbulence

$$\overline{i_{T}+i_{p}+d}=\int_{-\infty}^{\infty}T(U,r)P(U,r) dU=0\eqno(42)$$

\noindent
It will be shown below that (42) can serve
 as an equation for determination of
the only unknown parameter of the theory $B$. Comparing (40)-(41) 
with (22) gives:

$$-\frac{1}{2}\frac{\partial}{\partial U}T(U,r)P(U,r)=-\frac{u_{rms}}{L}\frac{\partial}{\partial U}UP(U,r)
 +(B^{o}-1)
\int_{-\infty}^{U} \frac{\partial P(y,r)}{\partial r}dy - A\frac{U}{r}P(U,r)\eqno(43)$$

\noindent
Recalling that $B=-B^{o}>0$ and $A=(3+B)/3>0$, the relations (42)-(43) give an equation
 for the only unknown coefficient $B>0$. 
It can be solved numerically to establish consistence with the above 
calculation leading to $B\approx 20$.
An interesting insight  into (42)-(43) is obtained if we use the fact  
that the deviations from 
scaling of a central part of the PDF are small. This means that:

$$P(U,r)=\frac{1}{r^{\alpha}}F(\frac{U}{r^{\alpha}})$$

Substituting this into (42)-(43) gives:

$$-\frac{1}{2}\frac{\partial }{\partial U} T(U,r)P(U,r)=-\frac{u_{rms}}{L}\frac{\partial }{\partial U}UP(U,r)
+\frac{\alpha(B+1)U}{r}P(U,r)-\frac{AU}{r}P(U,r)$$

\noindent Now, we have to define the ``representative exponent'' 
$\alpha\approx \xi_{n}/n$ with $0<n<3$, which can be done only approximately.
Choosing $\alpha =\xi_{1}$ and recalling that
 $\xi_{1}(B+1)\equiv A$,  the last two terms cancel  and 

$$T(U,r)=2\frac{u_{rms}}{L}U$$

\noindent satisfying the constraint (42).  

A much more interesting relation is obtained by substituting (36) into (43) with 
the scale-invariant expression for $P_{e}(U,r)$.  Taking into account that
 $\alpha (B+1)=A$, we have:

$$\frac{T(U,r)}{2}=\frac{2(B+1)(1-3\alpha)}{15P(U,r)}\frac{\partial P_{e}(U,r)}{\partial U}\eqno(44)$$

\noindent  The plot of $T(U,r)$
calculated from (39) using an approximate relation (36)-(37)
is presented on Fig. 7 for $r\approx 0.01$. 
 Outside the  interval   $-2~<U/r^{\alpha}<~3-4$  the curve
 saturates which might be an artifact of the exponential asymptotics   
the approximate formula (36)-(37),  valid only when $U$ not too large.
It is interesting that the curve is asymmetric, reflecting the asymmetry
 of the PDF.
In the limit $U\rightarrow 0$ the expression (44) gives:

$$T(U,r)\approx \frac{4(B+1)(3\alpha-1)}{15}\frac{U}{r^{2\alpha}}$$

\noindent This relation 
 takes into account that $3\alpha >1$. Choosing $\alpha\approx 
\frac{3+B}{3B}\approx \frac{23}{60}\approx 0.383$ gives:

$$T(U,r)\approx 0.84\frac{U}{r^{2\alpha}}\approx \frac{0.84 A_{2}}{r^{0.07}}
\frac{U}{S_{2}}\eqno(45)$$

\noindent where $A_{2}\approx 2$ is the amplitude of $S_{2}(r)$.

Coming back to the estimate of  parameter   $B$.  We saw that
 setting $\alpha=\xi_{1}$,
though  shedding  some light on the 
structure of the expression (43),  does not allow it's estimate. The problem 
is that $S_{1}=0$ and one cannot use it for the  non-dimensionalization of
the argument of the scale-invariant PDF.  A different parametrization 
  $\alpha=\xi_{2}/2$, 
typically used for analysis of experimental data,  
leads to
cancellation of the last two terms if

$$\frac{\xi_{2}}{2}({B+1})=\frac{B+3}{3}$$

\noindent giving $B\approx 40$. With the exponent
 $\alpha\approx 0.383$, best characterizing the top 
of the PDF $P_{e}(U,r)$ we have $B\approx 12$. This simple calculation 
demonstrates consistency of the model for $T(U)$ with the magnitude of the
parameter $B\approx 20$, derived above.

The expression (43) can be modified 
using an   identity [27]:

$$i_{1}(U,r)\equiv \frac{\partial P(U,r)}{\partial r}=
-\frac{\partial}{\partial U}<\frac{\partial U}{\partial r}|U>P(U,r)
\equiv\frac{\partial}{\partial U}I_{1}(U,r)P(U,r)$$

\noindent where $I_{1}(U,r)$ is the  conditional expectation value of 
$\frac{\partial U}{\partial r}$ for a fixed value of velocity difference $U$.  The 
equation (43) reads:

$$\frac{\partial}{\partial U}T(U,r)P(U,r)=-\frac{u_{rms}}{L}\frac{\partial}{\partial U}UP(U,r)
 +(B+1)I_{1}(U,r)P(U,r) - A\frac{U}{r}P(U,r)$$

It is easy to see that

$$\frac{1}{2}\frac{\partial U}{\partial r}=
\frac{u(r+\delta)-u(\delta)-u(r)+u(0)}{\delta}=u_{x}(r)-u_{x}(0)\eqno(46)$$

\noindent This expression shows  that experimental and numerical investigations of 
conditional 
expectation value of velocity-derivative difference for  a fixed value
 of $U$ is
  important since the combination

$$[(B+1)<\frac{\partial U}{\partial r}|U>-\frac{B+3}{3}\frac{U}{r}]P(U,r)
\eqno(47)$$

\noindent determines the structure of the probability density. Since
$P(U,r)=P(-U,-r)$,  we have from (35):

$$T(U,r)=-T(-U,-r)$$

\noindent{\bf Large-scale corrections to scaling}
\\
\noindent To complete comparison of the present work predictions 
with experimental data let us discuss some consequences of the
expression (26). Experimental determination  of the inertial range
is usually done  by establishing the interval where the 
third-order moment $S_{3}(r)\propto r$,
in accord with the Kolmogorov relation. As one can see from (26)
it is not so easy because of the   $O(rS_{2})$ subdominant
contribution giving

$$S_{3}\approx -0.8r+O( {r^{1.7}})$$

\noindent where we set $\xi_{2}\approx 0.7$ and $u_{rms}=L={\cal E}=1$.
To make quantitative comparison of  this relation with the data
we have to restore  physical dimensional units. Let us define

$$S_{n}(r)=A_{n}s_{n}\equiv A_{n}({\cal E}r)^{\frac{1}{3}}(\frac{r}{L})^{\xi_{n}-\frac{n}{3}}$$

The relation (26) can be rewritten:

$$\frac{S_{n}}{s_{n}}=A_{n}+\frac{u_{rms}}{({\cal E}L)^{\frac{1}{3}}}\frac{n(n-1)A_{n-1}}{(n+B)(\xi_{n-1}-\xi_{n}+1)}(\frac{r}{L})^{\xi_{n-1}-\xi_{n}+1}$$

 The high-Reynolds number experiment [23] was conducted in the 
atmospheric boundary layer on the tower at $35 m$ above the ground. 
The measured $u_{rms}\approx 1.4 m/sec$ and the mean dissipation rate 
${\cal E}\approx 0.03 \frac{m^{2}}{sec^{3}}$. The measured root-mean-square
stream-wise velocity component $u_{rms}$ in
the wall-bounded flows is somewhat larger than that  of the 
velocity components perpendicular to the direction of the flow. Since 
in the relation (26) we are 
interested in $u_{rms}$, corresponding to the top of
the inertial range, a good estimate for 
$u_{rms}/({\cal E}L)^{\frac{1}{3}}\approx 1$. Taking $L\approx 35 m$ and
$A_{2}\approx 2.0~-~2.5$ gives:

$$-\frac{S_{3}}{{\cal E}r}\approx 0.8~-~(0.06-0.08) r^{0.7}$$

\noindent and 

$$-\frac{S_{5}}{s_{5}}\approx 20~- ~2 r^{0.75}$$

\noindent
The experimentally observed [23] relation 

$$S_{5}=A_{5}({\cal E}r)^{\frac{5}{3}}(\frac{r}{L})^{\xi_{5}-\frac{5}{3}}
\approx 0.09r^{1.53}$$

\noindent 
is consistent with  numerical
values for ${\cal E}$ and $L$ used above. 
The third and fifth-order  moments,  calculated from the above relation, 
 are
 presented on Fig.8. The parameters used were: $A_{5}\approx 20$ and 
$A_{4}\approx 22$ [23]. The value of the integral scale $L\approx 35 m$, 
used above,
can be a bit overestimated. Choosing $L\approx 20-30~m~$  does not 
substantially modify
the above conclusions.
This result is extremely important since it shows that without explicit
 accounting
 for  the subdominant $O(r^{0.7})$- component of $S_{3}$, one
 cannot observe
the Kolmogorov
relation but in     very high Reynolds number flows. It also tells us that, in
 fact,  inertial
range can be made much broader and the scaling exponents can be established very accurately
with the proper data processing. One can see from the Fig.8 that the fifth-order moments
starts deviating from its asymptotic value earlier than $S_{3}$. The 
expressions corresponding to the model (23), can be written easily:

$$\frac{S_{2n+1}}{s_{2n+1}}=A_{2n+1}+
\frac{\gamma A_{2n-2}(2n+1)2n(2n-1)}{n+B}(\frac{r}{L})^{3+\xi_{2n-2}-\xi_{2n+1}}$$

\noindent As we see,  in the limit $r\rightarrow 0$ the contribution
from the sub-leading term in this case 
is much smaller. This means that for a given $Re$
investigation of the scaling exponents $\xi_{n}$ in the numerically- created 
large-scale-driven turbulence
is much easier.

\noindent{\bf Discussion and Conclusions}
\\

The theory developed in this work is based on  equations  including a simple
model for the pressure and dissipation terms. This model, though satisfying
all basic symmetry constraints, has not been   rigorously derived from the 
Navier-Stokes equations. The merits of such  work can be judged by 
comparison of theoretical predictions with experimental data. The calculated 
exponents and the amplitudes of the structure functions $S_{n}(r)$
 agree very well
with available experimental data. The theory also predicts the large-scale
 corrections to scaling, thus allowing calculation of $S_{n}$ up to
the large-scale cut- off $L$ 
at which $S_{2n+1}(r)$ become very small. 
This prediction is non-trivial and  can serve as a rigorous  
test of the model. The scale 
where it happens  is  an  integral scale of
 turbulence,
corresponding to the top of the inertial range. This definition seems 
very plausable since it corresponds to the length-scale of the non-zero 
energy flux set up. 

The most straitforward experimental test of this theory can be performed in a 
following way. Assuming that the odd-order moments have the form:

$$S_{2n+1}=A_{2n+1}r^{\xi_{2n+1}}+B_{2n+1}r^{\beta_{2n+1}}$$

\noindent where the exponents $\beta_{n}$, 
 reflecting the large-scale dynamics,
have been avaluated above for the two  cases of turbulence production.
The Log-Log plotting of the functions 

$$F_{2n+1}=B_{2n+1}r^{\beta_{n}-\beta_{2n+1}}
=-\frac{S_{2n+1}}{r^{\xi_{2n+1}}}-A_{n}$$

\noindent will enable one to obtain direct information about the sub-leading
contributions to the moments and, as a result, 
 directly assess the quality of the equations 
for the moments $S_{n}(r)$. The knowledge of $\beta_{n}$ will  define
universality classes,  differing by the mechanisms of turbulence
generation. According to this work, the functions $F_{2n+1}$ should 
demonstrate the scaling behaviour all the way up to the integral scale $L$. If
this is indeed so, the measurements are not too difficult.

If turbulence is driven by the large-scale body force, the subleading
correction to the Kolmogorov relation for $S_{3}(r)$ is an  analytic
$O(r^{3})$ function. This flow can be realized in numerical experiments.
In the real-life situations this kind of forcing rarely exist. 
The appearence of the $O(r^{1.7})$ non-analytic correction in the
 relation for $S_{3}(r)$,
derived from the equation (22),  can be easily explained in both 
 cases of decaying  and sheared
turbulence considered above. 
The measurements in jets and  wakes 
 are usually taken at 
the distance $x$ from
the origin (nozzles, bodies etc) where turbulence is produced. Thus, 
as was stated above,
the proper model is that  of decaying turbulence at the time $T=x/U$ after
turbulence generation  at $t=0$, where $U$ is the mean 
velocity at the crossection $x$. Then, assuming a close-to- self-similar decay
the correction is:

$$O(rS_{2t}(r))\approx rS_{2}(r)F_{t}(T)\approx r^{1.7}$$

\noindent where the function $F(t)$ describes the time-dependence 
of $S_{2}(r,t)$
in decaying turbulence. In case of the shear-generated turbulence
 the correction is [18]:

$$O(r \frac{\partial V}{\partial x} S_{2}(r))\approx r^{1.7}$$

\noindent  We do not know how general this result is.  According to 
present work,  it cannot be universal.  The theory  makes a direct
 connection to  Landau's remark about the role of the large-scale 
fluctuations of 
turbulence production. We cannot answer the most important question 
about universality of the exponents: for this we need detailed 
and quantitative information on the production terms in the equations 
of motion. However, even if the exponents are universal or belong
to some broad universality classes, the probability density function
of velocity difference is not universal since the amplitudes of the 
moments $S_{n}(r)$ depend on some of the features of the 
non-universal large-scale dynamics and the details of the single-point 
probability density. 

The assumption about  gaussian single point PDF, responsible for 
the values of the amplitudes of the even-order moments,  was used here as
a good approximation sufficient for demonstration of the basic features
of the theory. It is clear that the non-universality of the symmetric
part of the PDF $P_{e}(U,r)$ is determined by  deviations of the
 single-point PDF from the gaussian.
 It is interesting  
 that, according to the present work,  even neglecting non-universality
of $P_{e}(U,r)$, the asymmetric part $P_{o}(U,r)$, 
responsible for the odd-order moments, is not universal. 
This is quite reasonable since the very existence 
of the flux, reflected in $P_{o}(U,r)$, 
 is the result of the large-scale dynamics.

The theory,  presented here, is a departure from all previous 
field-theoretical 
attempts to develop an infra-red divergence-free  turbulence theory, 
 able to explain
 anomalous scaling of the moments of velocity difference.  It has  always
 been assumed that 
due to galilean invariance, the vertex corrections are equal to zero 
in the infra-red limit $k\rightarrow 0$. The supposed GI led to formulation
 of the Ward 
identities which were not too helpful.
The low- order Kraichnan's LHDIA [28],  which in addition to 
the conservation laws
correctly accounted for such  basic symmetries 
of the problem as random galilean invariance, 
led to the Kolmogorov energy spectrum without any corrections. It is
 interesting, that
 the same approximation, applied to Burgers
turbulence,  resulted in  the $k^{-2}$- energy spectrum corresponding to
 strong shocks. Kraichnan explained this 
non-trivial result in terms of the phase-decorrelation due to the interaction
 between components of the velocity field, non-existent in
 the Burgers dynamics,  
which effectively prevents the shock formation. According to [28],  it is this
decorrelation which is responsible for the formation of 
 the close- to- experimental data 
 Kolmogorov $5/3$-energy spectrum.
All  attempts to preserve  galilean invariance  of the theory 
 of three-dimensional turbulence led to
 disappearence of the integral scale $L$ from the problem and resulting 
inability 
of the theory to predict deviations from the Kolmogorov scaling. 
Polyakov's [11], Boldyrev [29] and Parisi et. al.[16] 
 theories of Burgers turbulence showed that
the galilean invariance is not to be taken for granted: only the low-order 
moments $S_{n\leq 1}\propto r^{n}$ corresponding to 
$|U|<<u_{rms}$ can be described in this regime.
The structure functions $S_{n}\propto r$ with $n>1$ scale with $u_{rms}$
 depending
on the large- scale features of the flow. These works stated  that
 galilean invariance is not sacred and  departures from it are responsible
 for the anomalous scaling of the high-order moments observed in Burgers 
turbulence. The same conclusion was derived in an  earlier work [30] 
predicting 

$$\overline{(u(x+r)-u(x)){\cal E}(x){\cal E}(x+r)}\propto u_{rms}
(\overline{{\cal E}})^{2}(r/L)^{0}$$

\noindent This experimentally confirmed relation [31]
explicitly involves  the GI breaking $u_{rms}$.

The present work makes an additional step in this direction: it assumes that 
due to incompressibility, GI  in  three-dimensional
turbulence is broken for all, even small velocity fluctuations. This means
that the ``normal scaling'' does not hold for all moments $n>-1$ 
which  seems to agree with experimental data showing deviations from 
Kolmogorov scaling  of all moments $S_{n}$. Free from the GI  restrictions, 
the vertex corrections are introduced in this paper from the very beginning  resulting 
in the equation of motion for the
probability density of velocity difference. This equation is  based on some general
symmetry properties of the system   and 
satisfies  all known realizability constraints. 

The theory developed here
 is based on a few assumptions. First of all, it assumes the existence of
 a  closed equation for  longitudinal structure functions $S_{n}(r)$.
 This is a mere generalization
 of the Kolmogorov result for a  particular case with $n=3$. 
Physical grounds for the equation for $S_{n}$
are based on the fact the Navier-Stokes non-linearities tend to produce two main effects:
the shock-generation  due to the advection terms which are balanced by the pressure contributions. An 
interplay between the two leads to creation of the vortical structures seen in the experimental data.
The three-dimensional nature of the structures is lost when one considers projection of the entire
dynamics onto a line. All we know is that the shock-production is effectively prevented by the pressure
 terms leading to invalidation of the bi-fractal description of the pure-Burgers dynamics.
We also know that the equation of motions are invariant under transformation $U\rightarrow -U$ and 
$x\rightarrow -x$ and that the dynamic equation for the $N$-point generation function must satisfy the 
general fusion rule transforming it, upon point merging, into the equation for the $N-1$-function.
These are  the physical reasons, responsible  for all,  but one,  contributions to the equation (20).

As was mentioned above, except for the
$A$-term in (20),  all others are more or less prescribed by the original
equation of motion. However, without the $A$-term, the equation 
(20) contradicts
exact relation (42) and thus cannot be correct. I have not been  
able to find an alternative description,
 obeying  general the fusion rules and symmetries  of
 the problem,  and producing solution satisfying  (42). For example, 
one can add 
the $O(Z_{2})$-contribution to the
 right side of (19) which does not contradict 
  the  basic symmetries. It violates, however, one of the principle constraints
 of the theory $S_{1}=<U>=0$ and
thus, cannot be correct. The success of the Boldyrev theory including this term in description
of  some of the regimes of  Burgers turbulence [28] is based on the fact that, as Polyakov's work,
it describes only the moments $S_{n}$ with $n<1$ and the result
$S_{1}=0$  can be achieved using   contributions coming from the non-scale invariant terms,  which
are beyond approximations of Refs.[11].[29]. The same happens in the Parisi et. al theory [16]
leading, in accord with the bi-fractal picture, to the PDF consisting of two contributions:
responsible for the moments with $n<1$ and $n>1$, respectively.  In the 
present paper we,  treating  all moments $S_{n}(r)$ with $n>0$ on
 equal footing,   do not have 
the luxury of satisfying the dynamical constraints using some contributions
extraneous to the theory 
and, as a result, our choice of the allowed terms in the equations of motion is much more narrow. One may also attempt to add 
 
$$h\frac{Z_{2}}{r\lambda}$$

\noindent not violating the symmetry of
the equation (20).   This gives 

$$\xi_{n}=\frac{n+h}{n+B}$$

\noindent  However, since $\xi_{0}=0$, the constant  $h=0$. 

The equation  (22) for the PDF can be rewritten in the limit of small
 $r$ as:

$$\frac{\partial P}{\partial t}+U\frac{\partial P}{\partial r}+
B^{o}\int_{-\infty}^{U}\frac{\partial P(y,r)}{\partial r}dy=\frac{P}{\tau}\eqno(48)$$

\noindent where $\tau\propto r/U$. 
The meaning of the 
$A$- term can be understood from  the shape  of (48) if we take into
 consideration that, unlike in the 
one-dimensional case, the interaction with the transverse components of the
 velocity field, produces an effective sources and  sinks or friction
for the longditudinal correlations. Then, (48) is the equation of motion
 taking these  sinks and sources 
 into account in the relaxation- time approximation. In other words
$\tau\approx r/U$ is simply the life-time (eddy turn-over time)
taking the longditudinal structure to substantially change its shape and size
due to 
interactions with the transverse components. This characteristic time, 
though plausible, 
is yet to be derived from the final ``microscopic theory''.
It is important that the equation (48)
is conservative, so  that,  since  $<U>=0$, $S_{0}=1$
. Thus,  the 
 right side of (48) describes both sinks and sources. This
 is consistent
with the results of the present paper: in the  small-scale limit: 
$S_{n}\propto r^{\xi_{n}}>>r^{\frac{n}{3}}\approx S_{nB}$ for $n<3$ while 
$S_{n}<<S_{nB}\propto r$ for all $n>3$. Here $S_{nB}$ is  the $n^{th}$-order
moment  of
 velocity difference measured in the large-scale-driven Burgers 
turbulence. This can be seen directly from the equation (48): the PDF
tail with  $U>0$  grows while the part those with $U<0$ decreases 
making the Navier-Stokes PDF $P(U,r)$ much more symmetric than the one governing
the Burgers dynamics. This means that due to the  
interaction  between
different components of the velocity field, 
the weak structures ($|U|<u_{rms}$),  generated by the 3D
 the Navier-Stokes dynamics are much more intense than their counterparts 
in the Burgers turbulence. At the same time,  due to Kraichnan's 
phase decorrelation, the strong structures in the Navier-Stokes turbulence
 are much weaker than the strong shocks, responsible for
 the high-order moments of the Burgers dynamics. 

In a recent paper by Zikanov et.al. [32] a modified one-dimensional 
Burgers equation

$$u_{t}+u_{x}u+\alpha u_{x}\int^{\infty}_{-\infty}\frac{u(x')}{x-x'}~dx'~=
~f+\nu u_{xx}\eqno(49)$$

\noindent has been considered. As in Ref. [12],  the white-in-time gaussian
random force characterized by the spectrum 
$\overline{|f(k)|^{2}}\propto k^{-1}$
was used. The large scale dissipation was introduced to avoid growth of 
the mode $u(k=0)$. It has been 
 shown that addition of the non-local contribution
is sufficient to prevent the shock formation and generation of the non-trivial
exponents $\xi_{n}\neq 1$ for $n>3$. The possible relation of (49) to 
the equations, introduced in this paper, will be discussed elsewhere.  

The good agreement  of the scaling exponents of the structure functions  
wiht experimental data, derived in this paper,  though  gratifying, is 
not the most important outcome. The cascade models, not related to the equations
 of motion, 
also give  quantitatively correct $\xi_{2n}$. However, no model was able
 to address the 
problem of the asymmetry of the probability density function
 $P(U,r)\neq P(-U,r)$ and, as a consequence,
predict the scaling exponents and the amplitudes of the odd-order structure functions.
Only dynamic theory,  based on  
the Navier-Stokes equations, 
invariant under
transformation $u\rightarrow -u$ and $x\rightarrow -x$ can lead to the 
asymmetric probability density and correct properties of the odd-order moments.

The present theory, though based on some physical considerations  developed for
 the Burgers equation, is not a small
perturbation around  Burgers phenomenology: the coefficient $B\approx 20$
and $B>>b\approx 2$ obtained in [11]. Moreover, the relevant coefficient
 $B^{0}$, renormalizing 
advection contributions in equation (19), is strongly negative unlike
parameter $b>0$ in the theory of Burgers turbulence. 
This means that  the pressure and transverse terms,  preventing 
formation of strong shocks, 
are extremely important here.
 On the other hand,
the Burgers effects are not unimportant at all.
To demonstrate it,  we can neglect
the ``original''  Burgers terms in the equation of motion and derive
the relation  for the moments:

$$S_{n}=A_{n}r^{\kappa_{n}}-
\frac{u_{rms}}{L}\frac{A_{n-1}}{A_{n}}n(n-1)
\frac{r^{\kappa_{n-1}+1}}{1-\frac{A}{B'}}$$

\noindent where now
 
$$\kappa_{n}=\frac{A}{B'}n$$

\noindent and $A/B'<1$. We see that without ``small'' Burgers terms the equation
 of motion
gives normal scaling and that is why they are essential for derivation of the 
anomalous scaling exponents. If all this is correct, one  may say that the anomalous
 scaling is the result 
of the dynamic interplay of the Burgers-like tendency to create singularities (shocks) with 
the ``normalizing'' action of the pressure terms and the incompressibility constraints.

The equations 
developed here are based on the phenomenology relevant for the
 longitudinal structure functions and that is why we cannot say anything 
about  shape and scaling of transverse structure functions. 
The main problem is that 
the odd-order  moments $S^{t}_{2n+1}=
<(v_{\alpha}({\bf x+r})-v_{\alpha}(\bf x))^{2n+1}>=0$,  where vector ${\bf r}$
is parallel  and  $v_{\alpha}$ is a component of the velocity
 field perpendicular to the $x$-axis. This  means that the PDF 
$P(\Delta v_{\alpha}, r)=P(-\Delta v_{\alpha},r)$ and the equations  governing 
 probability density of transverse velocity differences must have 
 different
symmetry properties than (20)-(21).

The most important feature of hydrodynamic turbulence, distinguishing it 
from equilibrium statistical mechanics, is constant energy flux in the wave-number space.
It is this energy flux that makes the probability density $P(U,r)$ asymmetric,  leading to the 
non-zero values of the odd-order longitudinal structure functions. It 
is not clear how the information
about the energy flux is reflected in the transverse structure functions coming out
from the corresponding symmetric probability density. It is even 
unclear if $S^{t}_{n}(r)$ are  a dynamically relevant object. In the theory 
presented here, the transverse components of the velocity field simply
serve as 
a ``bath'' introducing some renormalization and dephasing into the 
 the energy flux- carrying longitudinal dynamics. The angular dependence 
of the structure functions in three-dimensional turbulence can be recovered
using the multidimensional equation for the two-point generating
 function  with the pressure terms accounted for in the mean field
approximation,  similar to the one 
introduced in this paper. It is not clear if this  can lead to 
the improved description of experimental data.

\noindent{\bf Acknowledgments}
I am grateful to R.H. Kraichnan whose remarks, insights  and constructive 
suggestions were
most essential for this work. My thanks are due to S.-Y.Chen,
K.R.Sreenivasan, B. Dhruva, R.Miles and U.Frisch 
 for providing me with their most recent,
sometimes unpublished, experimental data. Very interesting and stimulating 
discussions with S. Boldyrev, A. Chekhlov, 
M. Chertkov, U. Frisch,  M. Nelkin, A.Polyakov and B. Shraiman 
are gratefully acknowledged. This
work was supported in part by the ONR/URI grants.

\noindent {\bf references}
\\
1. L.D.Landau and E.M. Lifshitz, Fluid Mechanics, Pergamon Press, Oxford, 1987;
~~A.S.Monin and A.M.Yaglom, ``Statistical Fluid Mechanics'' vol. 1, MIT Press,
Cambridge, MA (1971)
\\
2. A.N. Kolmogorov, J.Fluid Mech., {\bf 5}, 497 (1962). ~~ An interesting 
 chain of events 
 starting 
with Landau's remark and ending with the Kolmogorov paper is described
in:
U. Frisch, ``Turbulence'', Cambridge University Press, 1995. 
\\
3. R.H. Kraichnan, Phys.Rev.Lett.,{\bf 72}, 1016 (1994)
\\
4. F. Gawedzki and A. Kupianen, Phy.Rev.Lett., {\bf 75}, 3834 (1995)
\\
5.  M. Chertkov, G. Falkovich, I. Kolokolov and V.Lebedev, Phys.Rev. E,{\bf 52}, 4924 (1995)
\\
6. M. Chertkov, G. Falkovich, I. Kolokolov and V.Lebedev, Phys.Rev. E,{\bf 51}, 5609 (1995).
\\
7. B. Shraiman and E. Siggia, CR Acad.Sci., {\bf 321}, Serie II (1995)
\\
8. V.Yakhot, Phys. Rev. E,       (1996)
\\
9. M.Vergassola and A.Mazzino, CHAO-DYN/9702014
\\
10. O.Gat, V. L'vov, E. Podivilov and I. Procaccia CHAO-DYN/9610016 
\\
11. A.M. Polyakov, Phys.Rev. E {\bf 52}, 6183 (1995)
\\
12. A. Chekhlov and V. Yakhot, Phys.Rev.E {\bf 51}, R2739 (1995)
\\
13. A. Chekhlov and V. Yakhot, Phys.Rev. E {\bf 52}, 5681 (1995)
\\
14. V. Yakhot and A.Chekhlov, Phys.Rev.Lett. {\bf 77}, 3118 (1996)
\\
15. M. Chertkov, I. Kolokolov and M. Vergassola, preprint
\\
16. J.P. Bouchaud, M. Mezard and G. Parisi, Phys.Rev. E {\bf 52} 3656 (1995);
~V.Gurarie and A.Migdal, Phys. Rev. E{\bf 54},4908 (1996);~~the relation for 
$S_{4}(r,\theta)$ has been written by M.Nelkin;
~~A.Chekhlov and V.Yakhot, unpublished (1996)
\\
17. D. Forster, D. Nelson and M. Stephen, Phys. Rev. A{\bf 16} 732 (1977)
\\
18. S. Corrsin, NACA R$\&$M 58B1 (1958);
 J.L. Lumley, Phys. Fluids {\bf 10}, 855, (1967);
I am grateful to B. Shraiman who brought my attention to these  works.
\\
19. S.-Y. Chen, private communication, 1997
\\
20. M. Chertkov, Phys.Rev. E, {\bf 55}, 2722  (1997)
\\
21. P. Kailasnath, A.A. Migdal, K.R. Sreenivasan, V. Yakhot and L. Zubair,
Yale preprint (1991)
\\
22. S.-Y. Chen, private communication, 1997
\\
23. B. Dhruva and K.R. Sreenivasan, private communication (1997)
\\
24. A. Noullez,
 G. Wallace, W. Lempert,   R.B. Miles and U.Frisch, J.Fluid.Mech. 
{\bf 339}, 287 (1997)
\\
25. Ya. G. Sinai and V.Yakhot, Phys.Rev.Lett., {\bf 63},1962 (1989)
\\
26. S. Pope, Combust. Flame {\bf 27}, 299 (1976)
\\
27. I am gratefull to R.H.Kraichnan who pointed out this relation to me
\\
28. R. H. Kraichnan, Phys.Fluids. {\bf 11}, 266  (1968)
\\
29. S. Boldyrev, Phys.Rev.E {\bf 55}, 6907 (1997)
\\
30. V. Yakhot, Phys.rev. E {\bf 50}, 20 (1994)
\\
31. A. Praskovskii and S. Oncley, Phys. Rev. E {\bf 51}, R5197  (1995)
\\
32. O. Zikanov, A. Thess and R. Grauer, Phys.Fluids {\bf 9},  1362 (1997)

\newpage

{\bf Figure legends}
\\
\noindent Fig. 1. Comparison of the calculated scaling exponents (formula (27))
with the results of numerical simulations of Chen [19]. 
\\
\noindent Fig. 2. Probability density of 
velocity differences 
$F(x)=r^{\frac{1}{3}}P(\frac{U}{r^{\frac{1}{3}}})$ vs $x=\frac{U}{r^{\frac{1}{3}}}$. 
\\
\noindent Fig. 3. Probability density $P(U,r)$ as a function of $U/u_{rms}$.
\\
\noindent Fig. 4. Mesured PDF's of transverse structure functions [24]
for a few values of the displacement $r=3600 \mu m;~ 900 \mu m;~ 28 \mu m;$.
\\
\noindent Fig. 5. Approximate parametrization of the PDF $r^{\alpha}P(U,r)$
 vs $x=\frac{U}{r^{\alpha}}$ using formula (37).
\\
\noindent Fig. 6. Asymmetric PDF  $P(U,r)$ given by (36)-(37).
\\
\noindent Fig. 7. Calculated conditional mean $\frac{T(U,r)}{2}$ using approximate
expression for the PDF (36)-(37).
\\
Fig. 8. Calculated normalized moments $s_{3}=\frac{S_{3}(r)}{r}$ and 
$s_{5}=\frac{S_{5}(r)}{r^{1.53}}$.

\newpage

\begin{figure}[h]
\centerline{\psfig{file=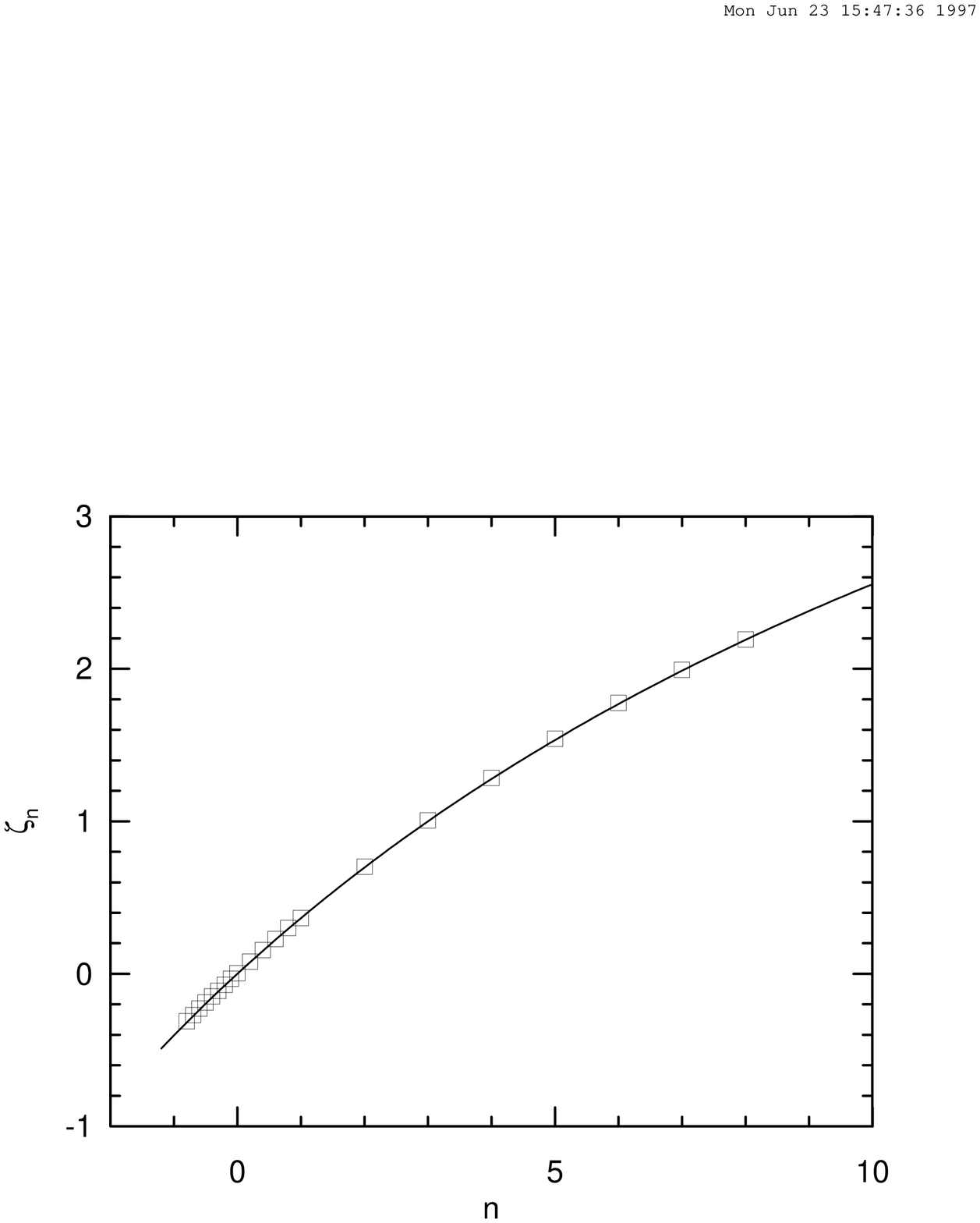,width=5.5in,height=5.0in}} 
\caption{Comparison of the calculated scaling exponents (formula (27))
with the results of numerical simulations of Chen [19]} 
\end{figure}

\begin{figure}[h]
\centerline{\psfig{file=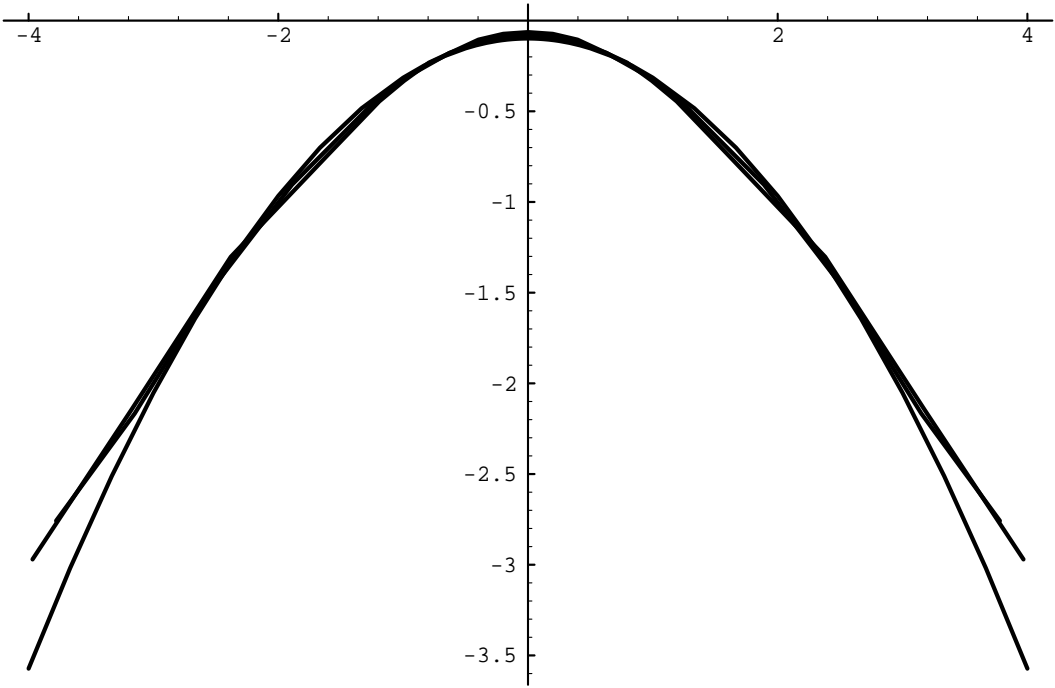,width=5.5in,height=8.0in}} 
\caption{Probability density of 
velocity differences 
$F(x)=r^{\frac{1}{3}}P(\frac{U}{r^{\frac{1}{3}}})$ vs $x=\frac{U}{r^{\frac{1}{3}}}$} 
\end{figure}

\begin{figure}[h]
\centerline{\psfig{file=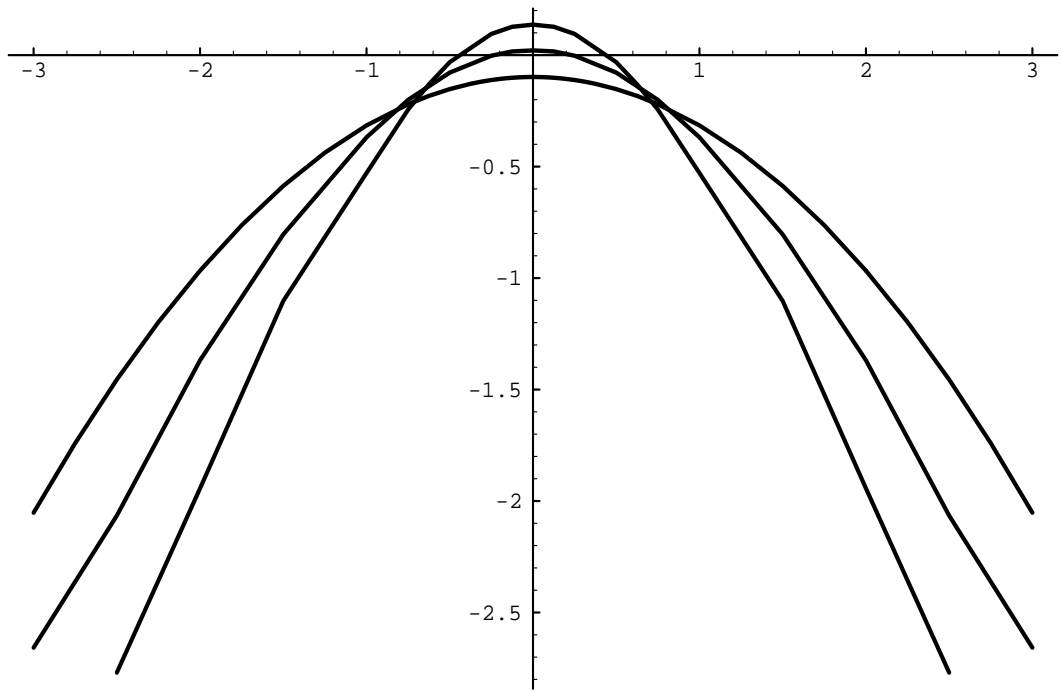,width=5.5in,height=8.0in}} 
\caption{Probability density P(U,r) as a function of U} 
\end{figure}

\begin{figure}[h]
\centerline{\psfig{file=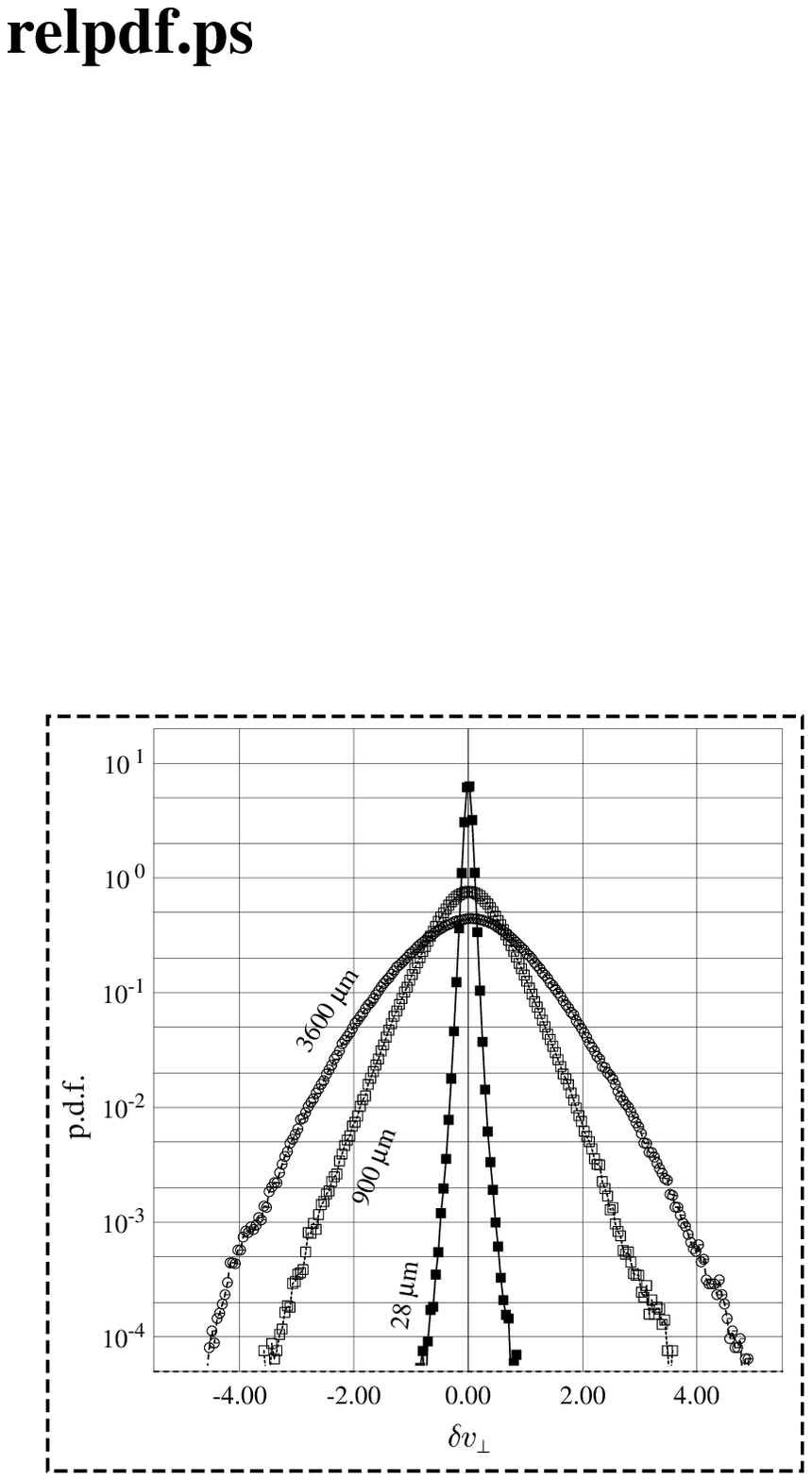,width=5.5in,height=5.0in}} 
\caption{Mesured PDF's of transverse structure functions [24] } 
\end{figure}

\begin{figure}[h]
\centerline{\psfig{file=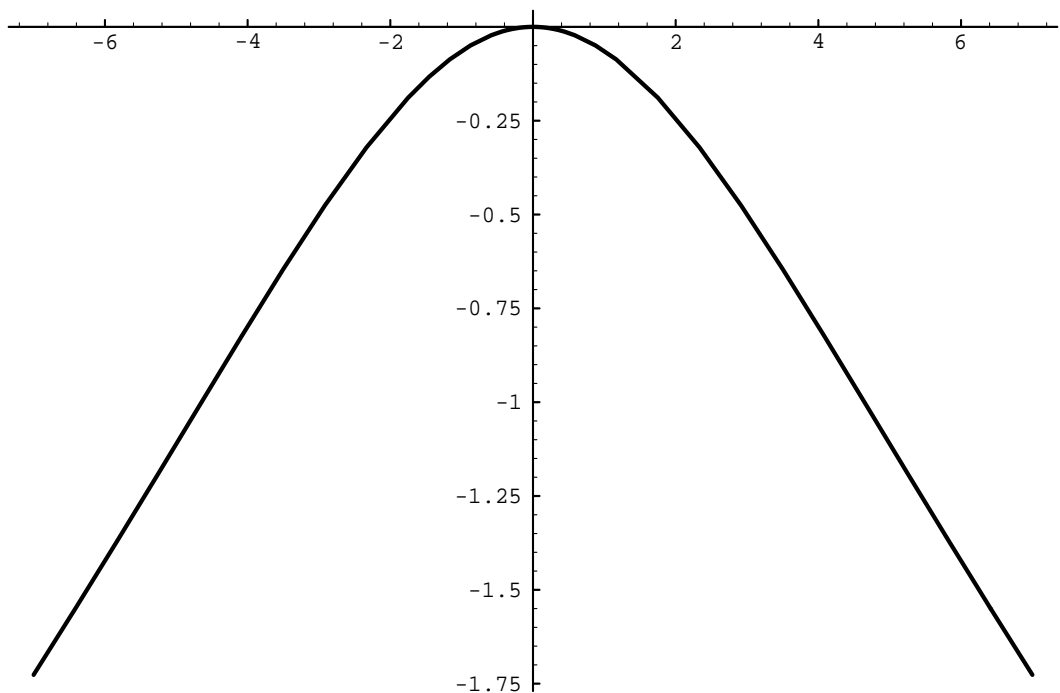,width=5.5in,height=8.0in}} 
\caption{Approximate parametrization of the PDF $r^{\alpha}P(U,r)$
 as a vs $x=\frac{U}{r^{\alpha}}$ using formula (37) } 
\end{figure}

\begin{figure}[h]
\centerline{\psfig{file=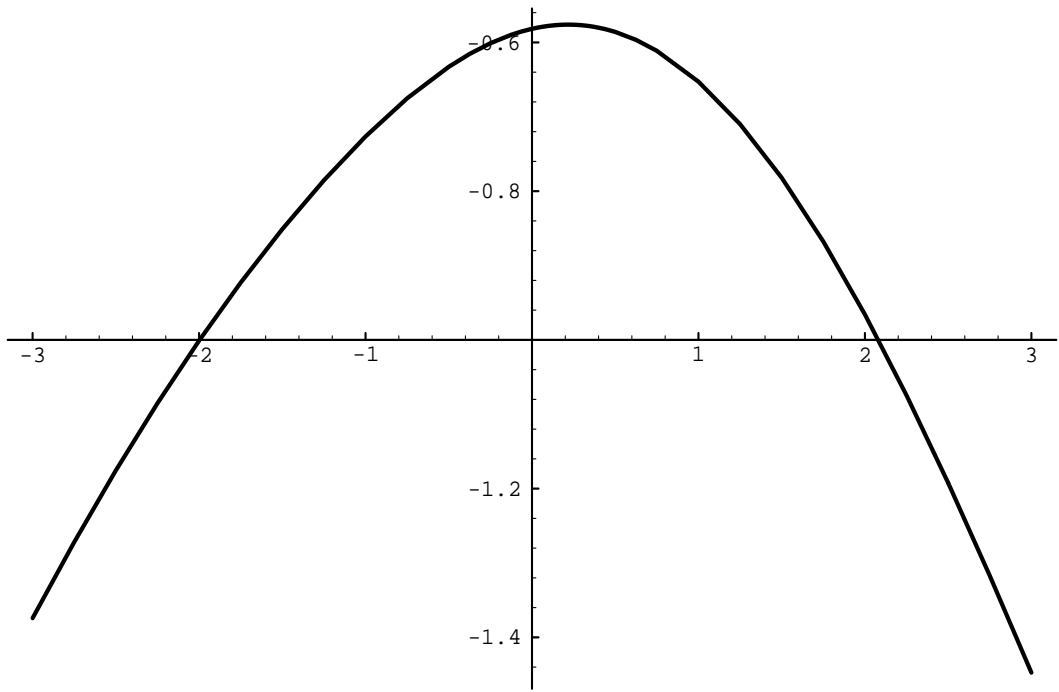,width=5.5in,height=8.0in}} 
\caption{Asymmetric PDF  P(U,r) given by (31)-(32) } 
\end{figure}

\begin{figure}[h]
\centerline{\psfig{file=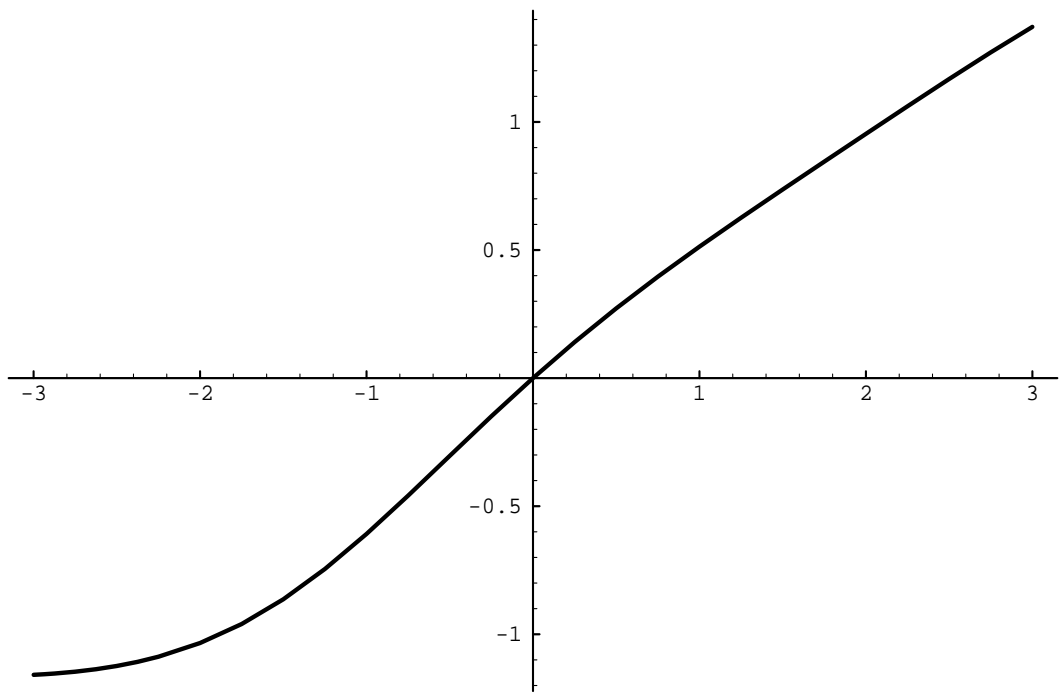,width=5.5in,height=8.0in}} 
\caption{Calculated conditional mean $\frac{T(U,r)}{2}$ using approximate
expression for the PDF (36)-(37) } 
\end{figure}

\begin{figure}[h]
\centerline{\psfig{file=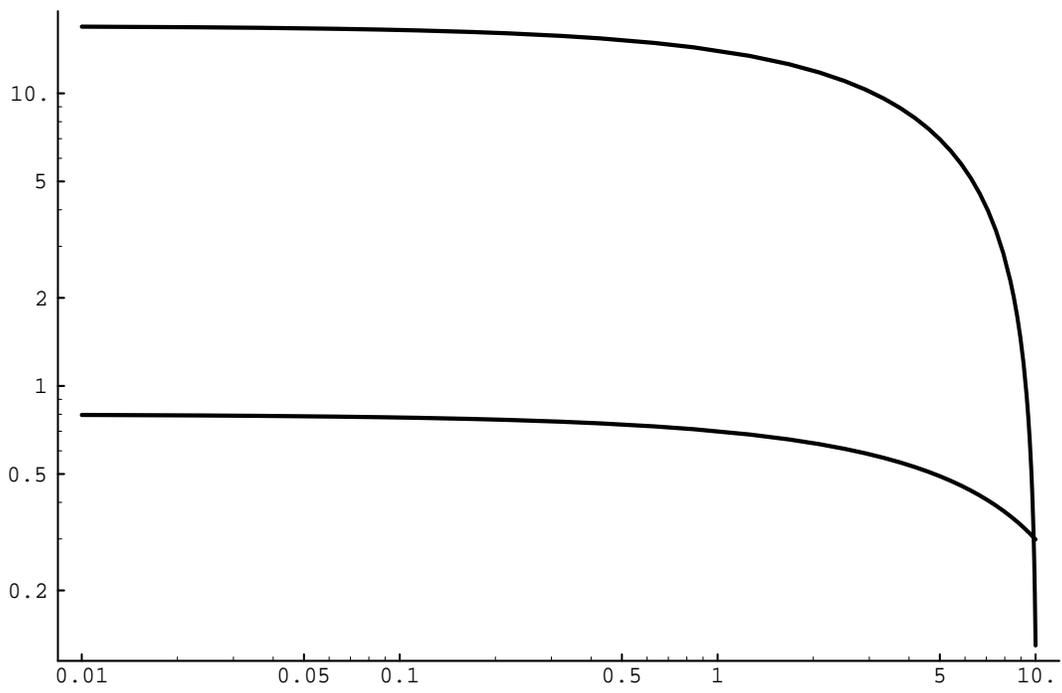,width=5.5in,height=8.0in}} 
\caption{Calculated normalized moments $S_{3}(r)$ and $S_{5}(r)$ } 
\end{figure}

\end{document}